\documentclass[12pt]{article}

\usepackage[dvipsnames]{xcolor}
\usepackage{enumerate,enumitem}

\usepackage{amsmath,amssymb,amsfonts,mathrsfs, amsthm,mathtools, bm, bbm, dsfont, mathrsfs, amsthm}
\usepackage{graphicx, epstopdf}

\usepackage{fullpage}

\usepackage{ebgaramond}
\usepackage[OT1]{fontenc}
\usepackage[utf8]{inputenc}

\usepackage[colorlinks=true,breaklinks=true,bookmarks=true,urlcolor=MidnightBlue,citecolor=MidnightBlue,linkcolor=MidnightBlue,bookmarksopen=false,draft=false]{hyperref}

\allowdisplaybreaks

\newcommand{\qedwhite}{\hfill \ensuremath{\Box}}

\newcommand{\CS}{{{{\sf OCS}}}}
\newcommand{\PCS}{{{{\sf OPCS}}}}
\newcommand{\NCS}{{{{\sf OS}}}}
\newcommand{\SYM}{{{{\sf SYM}}}}
\newcommand{\SYMA}{{{{\sf SYMA}}}}
\newcommand{\PI}{{{{\sf PI}}}}

\usepackage{booktabs} 

\newtheorem{theorem}{Theorem}[]
\newtheorem{proposition}{Proposition}[]
\theoremstyle{definition}
\newtheorem{definition}{Definition}[]

\theoremstyle{assumption}
\newtheorem{assumption}{Assumption}[]

\theoremstyle{remark}
\newtheorem*{remark}{Remark}
\begin{document}

\title{Incentive Designs for Stackelberg Games with a Large Number of Followers and their Mean-Field Limits}

\author{Sina Sanjari \quad Subhonmesh Bose \quad Tamer Ba\c{s}ar\thanks{Sina Sanjari is with the Department of Mathematics and Computer Science at the Royal Military College of Canada. Subhonmesh Bose and Tamer Ba\c{s}ar are with the Department of Electrical and Computer Engineering at the 
     University of Illinois Urbana-Champaign (UIUC). T. Ba\c{s}ar is additionally with the Coordinated Science Laboratory at UIUC. Emails: sanjari@rmc.ca and \{boses, basar$1$\}@illinois.edu. This research was supported in part by grant FA9550-19-1-0353 from AFOSR, and in part by the National Science Foundation under the grant CPS-2038403.}}

\maketitle

\begin{abstract}
We study incentive designs for a class of stochastic Stackelberg games with one leader and a large number of (finite as well as infinite population of)  followers. We investigate whether the leader can craft a strategy under a dynamic information structure that induces a desired behavior among the followers. For the finite population setting, under convexity of the leader's cost and other sufficient conditions, we show that there exist symmetric \emph{incentive} strategies for the leader that attain approximately optimal performance from the leader's viewpoint and lead to an approximate symmetric (pure) Nash best response among the followers. Leveraging functional analytic tools, we further show that there exists a symmetric incentive strategy, which is affine in the dynamic part of the leader's information, comprising partial information on the actions taken by the followers. Driving the follower population to infinity, we arrive at the interesting result that in this infinite-population regime the leader cannot design a smooth ``finite-energy''  incentive strategy, namely, a mean-field limit for such games is not well-defined. As a way around this, we introduce a class of stochastic Stackelberg games with a leader, a major follower, and a finite or infinite population of minor followers, where the leader provides an incentive only for the major follower, who in turn influences the rest of the followers through her strategy. For this class of problems, we are able to establish the existence of an incentive strategy with finitely many minor followers. We also show that if the leader's strategy with finitely many minor followers converges as their population size grows, then the limit defines an incentive strategy for the corresponding mean-field Stackelberg game. Examples of quadratic Gaussian games are provided to illustrate both positive and negative results.  In addition, as a byproduct of our analysis, we establish existence of a randomized incentive strategy for the class mean-field Stackelberg games, which in turn provides an approximation for an incentive strategy of the corresponding finite population Stackelberg game. 
\end{abstract}

\section{Introduction.}
Incentive design problems are hierarchical decision-making problems that involve a leader and a collection of followers with possibly different goals. The leader moves first and announces a strategy that is imposed on the followers. Then, the followers act simultaneously, following which the leader implements her strategy. In this setup, the leader's strategy is such that she has access to the followers' actions (either individually or a function thereof), and as a result, the followers' costs become functions of their collective actions and the leader's announced strategy. Stochastic incentive design problems add an extra layer of richness in that each player can have access to private information, that other players (except the leader) may not be privy to. In addition, the costs of all players might depend on random states of nature that remain unknown to all players. Stochastic incentive design problems therefore involve hierarchical decentralized decision-making with dynamic information structures.

The focus of study in all these games is the design of a leader's strategy that induces a desired behavior among the followers. In other words, the leader seeks to shape the collective (Nash) response of the followers toward her desired, possibly social, goal. As a specific example, consider the problem of designing the federal tax code. The government (leader) chooses a taxation policy. The citizens file their tax returns. The government collects taxes based on the announced taxation policy and the returns filed by the citizens. In this setup, one seeks a taxation policy that induces socially equitable tax payments from the citizens. Another example is a duopoly with government regulation. The government can act as the leader and design regulation strategies to induce competitive behavior among the duopolistic firms (the followers) towards a Pareto-optimal solution \cite{salman1981incentive}. See  \cite{zheng1984stackelberg, ratliff2019perspective} for a variety of applications of incentive design problems. 

As one extreme case of the interaction between the leader and the followers, we can consider the scenario where the leader enforces followers to choose a particular set of actions/strategies by announcing a \emph{threat} policy--one that heavily penalizes each follower for even minutely deviating from response desired by the leader. While effective, such strategies are unsuitable for policy-makers to implement. Ideally, the loss or penalty incurred by each follower for a deviation from a desired response should be smooth and commensurate with the extent of that deviation; we pursue the design of such incentive strategies in this paper. Namely, we study a class of stochastic incentive design problems through a stochastic Stackelberg game formulation with decentralized information structure, and we investigate the existence and properties of smooth incentive strategies. Incentive design problems are somewhat different in spirit from mechanism design (e.g., in \cite{groves1973incentives, fudenberg1991game, groves1979incentives, dasgupta1979implementation}) in that the leader cannot change the nature of the interaction to attain the desired performance level, but rather shapes the followers' responses through its design of an incentive strategy alone.



The study of incentive design problems has a long history, e.g., see \cite{bacsar1984affine, cansever1985stochastic, bacsar1983performance, bas89, basols99, ho1982control, zheng1982existence, zheng1984stackelberg}. Here, we study incentive design within the framework of stochastic Stackelberg games with one leader and many (finite or countably infinite) number of followers. In Stackelberg games, when the leader does not have dynamic information involving the actions of the followers, one can often directly characterize equilibria, e.g., in \cite[Chapter 7]{basols99}. With dynamic information for the leader, such a direct approach to equilibrium characterization is often untenable. Rather, one can avail an indirect route and utilize the leader's desired strategy profile for the entire population to construct an incentive strategy for the game that sustains a Stackelberg equilibrium with the same leader's performance as the leader's optimal strategy \cite{bacsar1984affine}. In this work, we take this indirect route also but delineate the difficulties of transporting earlier results to our setting. There are three main challenges for solving stochastic incentive design problems with many followers. When multiple followers are present, we must undertake the challenging task of studying Nash best response(s) of the followers to the admissible strategies of the leader. Such responses may fail to exist and can even be non-unique—a setting strikingly different from requiring an optimal response from a single follower as in \cite{bacsar1984affine}. Furthermore, in the multi-follower setting, the indirect approach requires the solution of the leader’s optimal strategy which depends on the followers’ actions. This requires the solution of a decentralized control problem with dynamic information that is inherently more complex than the static information counterparts studied in \cite{basols99, simaan1973stackelberg, simaan1973stackelberga}. Third, the stochastic variant of the incentive design problems requires optimization over infinite-dimensional strategy spaces, where each strategy is a measurable map of the available information, and hence, requires functional analytic and probabilistic tools. Such difficulties do not arise in deterministic incentive design problems, as analyzed in \cite{ bacsar1983performance, ho1982control, zheng1984stackelberg, zheng1982existence, tolwinski1981closed}. 
    
Building on our study of the problem with finitely many followers, we then seek an extension of the setup to that with an infinite number of followers, i.e., we study its mean-field limit. Our study of such limits is inspired by the rich and growing literature on mean-field games, e.g., see \cite{LyonsMeanField, CainesMeanField1, CainesMeanField2} for the early results on non-hierarchical mean-field Nash games. Our work adds to the nascent literature on mean-field Stackelberg games in \cite{moon2018linear, moon2016discrete, bensoussan2015mean, vasal2022master, moon2020linear,  yang2021linear, mukaidani2021robust}. These papers generally adopt the direct approach of characterizing Stackelberg equilibria, focusing mostly on linear quadratic Gaussian settings, those under the classical information structures and/or open-loop with open-loop policies that depend only on the history of disturbances and not history of states and/or actions. Our models and results in this paper go beyond quadratic Gaussian games, we consider Stackelberg games with decentralized dynamic information structures, whose intricacies render the direct approach to equilibrium characterization, implausible. Recall that our approach to equilibrium is indirect; we construct an incentive strategy, starting from an optimal or near-optimal strategy for the leader and study its mean-field limit. To our knowledge, this is the first paper on mean-field Stackelberg games that takes an indirect approach to equilibrium construction and even the question of when such equilibria exist. 
    
\subsection{Contributions.}

\begin{enumerate}
\item {For a general class of dynamic symmetric stochastic Stackelberg games with a finite number of followers, we establish the existence of a Stackelberg equilibrium. Our approach is indirect and uses the notions of incentive strategy and leader-optimality. To this end, we first show in Theorem \ref{the:symmetric} that there exists an approximate continuous symmetric leader-optimal strategy (such that from the leader's vantage point entails full cooperation of the followers) that leads to approximate symmetric Nash best response strategies for the followers. Then, we prove in Theorem \ref{the:sym2} that there exists a Stackelberg equilibrium that is symmetric and of incentive type such that the leader, together with a symmetric Nash best response of followers, approximately attains the leader-optimal performance.}

\item When the follower population is driven to infinity, we show in Proposition \ref{the:neg1} that no smooth incentive strategy exists for a quadratic Gaussian game with finite energy, i.e., such incentive design games may not admit a well-defined mean-field limit.

\item We introduce a class of stochastic Stackelberg games with one leader, one major follower, and finite and/or infinite number of minor followers, where the leader only incentivizes the major player. In such games, the power of the leader is limited, and cannot generally attain the performance that she can get by incentivizing the entire population of followers.
We show through a quadratic Gaussian game example that such games admit an incentive strategy that leads to a well-defined mean-field limit. 
In the general case, we establish in Theorem \ref{the:Ma-Mi} that if the sequence of incentive strategies for the finite population setting converges, then the limiting strategy defines an incentive strategy for the corresponding mean-field Stackelberg game. 

\item For the class of stochastic Stackelberg games with a leader, one major follower, and a finite number of minor followers, in Proposition \ref{lem:exi-1}, we establish approximations for Stackelberg-incentive equilibria within \emph{randomized} strategies by approximate Stackelberg-incentive equilibria that admit pure strategies for the leader and the major follower (not necessarily for the minor followers). Using this result, we then prove the existence of an approximate mean-field incentive equilibrium within randomized strategies in Theorem \ref{the:Existence}(i) when the minor population size is driven to infinity.

\item As a byproduct, we provide approximation results for symmetric incentive strategies for the game with a large but finite number of minor followers, using a mean-field incentive equilibrium for the limiting mean-field game in Theorem \ref{the:Existence}(ii). 

\end{enumerate}


\section{A Symmetric Stochastic Stackelberg Game with One leader and A Finite Number of Followers $\mathcal{P}_N$.}
\label{sec:sg}

We study a single-stage Stackelberg game $\mathcal{P}_N$ with dynamic information structure, where we identify the leader as player $0$, and the followers as players $1, \ldots, N$. In this game, the leader announces a strategy in the beginning. Then, the $N$ followers act simultaneously. However, the realized costs of the leader and the followers depend on the leader's announced strategy as well as the actions taken by the followers. In this section, we formally define this game as $\mathcal{P}_N$ and describe relevant equilibrium/optimality notions that we study in the sequel.

Let $(\Omega, \mathcal{F}, \mathbb{P})$ be the underlying probability space describing the system's distinguishable events. Let $\mathbb{Y}^0$ be a subset of a finite-dimensional Euclidean space, endowed with its Borel $\sigma$-field $\mathcal{Y}^0$ that describes the possible private observation $y^0$ of the leader. Let $(\mathbb{Y}, \mathcal{Y})$ describe the same for each follower. 
Also, let $\mathbb{U}^0$ be a subset of a finite-dimensional Euclidean space, that together with the Borel $\sigma$-field $\mathcal{U}^0$, describe the space of control actions $u^0$ for the leader. Similarly, define $(\mathbb{U}, \mathcal{U})$ and $u^i$ for each follower $i=1,\ldots, N$. Note that the action sets of different followers are identical.

Each player selects a control action via an admissible strategy--a measurable map of her available information. Followers are only privy to their private static observation, i.e., $I^i = \{y^i\}$ is the information available to follower $i$. Let $\Gamma^i$ denote her set of admissible strategies-- a set of measurable functions $\gamma^i$ from $(\mathbb{Y}, \mathcal{Y})$ to $(\mathbb{U}, \mathcal{U})$. For the leader, we consider the following three information structures.
\begin{itemize}
\item  Observation- and control-sharing: $I^0_{\CS}:=\{y^{0}, , y^{1}, \dots, y^{N}, u^{1}, \dots, u^{N}\}$.

\item { Observation and partial control-sharing: 
$I^0_{\PCS}:=\{y^{0}, y^{1}, \dots, y^{N}, \hat{\Lambda}(u^{1}, \dots, u^{N})\}$,
where $\hat{\Lambda}:\prod_{i=1}^{N}\mathbb{U}\to \mathbb{U}$ is a measurable function of the control actions of all followers.}

\item Observation-sharing: $I^0_{\NCS}:=\{y^{0}, y^{1}, \dots, y^{N}\}$.

\end{itemize} 
When only observations are shared, the leader's information and hence, her actions cannot adapt to the followers' actions. Hence, the leader's strategies with observation-sharing can be viewed as ``open-loop''. When the control actions of followers are shared, either individually or partially, the leader's policies are ``closed-loop''.
Let $\Gamma^0_\CS$, $\Gamma^0_\PCS$ and $\Gamma^0_\NCS$ denote the set of admissible strategies for the leader under the corresponding information structures, that is each of these sets include measurable functions $\gamma^0$ from her information available in the corresponding information set to $\mathbb{U}^0$.

Let $\omega_{0}$ be an $\Omega_{0}$-valued random variable that defines the common exogenous uncertainty that affects all players' observations and/or costs. Each player seeks to minimize her expected cost, given by
\begin{flalign}
J_{N}^{0}({\gamma}^{0:N}) 
&= \mathbb{E}^{{\gamma}^{0:N}}[c^{0}(\omega_{0},{u}^{0}, \Lambda(u^{1},\dots, u^{N}))], 
\label{eq:sym1.1}
\\
J_{N}^{i}({\gamma}^{0:N}) 
&= \mathbb{E}^{{\gamma}^{0:N}}\left[c(\omega_{0},{u}^{0}, {u}^{i}, \Lambda(u^{1},\dots, u^{N}))\right], 
\quad i = 1,\ldots, N
\label{eq:sym1.2},
\end{flalign}
for Borel-measurable $c^{0}: \Omega_{0} \times \mathbb{U}^{0} \times \mathbb{U} \to \mathbb{R}_{+}$ and $c: \Omega_{0} \times \mathbb{U}^{0} \times \mathbb{U} \times \mathbb{U} \to \mathbb{R}_{+}$. Here, we use the notation $\gamma^{j:k}:= \left(\gamma^j, \gamma^{j+1}, \ldots, \gamma^k\right)$ and $\mathbb{E}^{{\gamma}^{0:N}}$ denotes the expectation with respect to $\mathbb{P}$ when actions of players are written as measurable functions ${{\gamma}^{0:N}}$ of their observations.

We assume that a measurable function $\Lambda:\prod_{i=1}^{N}\mathbb{U} \to \mathbb{U}$  is permutation-invariant, i.e.,
\begin{flalign}
\Lambda(u^{1},\dots, u^{N})=\Lambda(u^{\sigma(1)},\dots, u^{\sigma(N)})
\label{eq:lambda}
\end{flalign}
for a permutation $\sigma$ of the set $\{1,\ldots,N\}$, which makes \eqref{eq:sym1.2} symmetric across all followers. {A specific example is the mean-field interaction case, where $\Lambda(u^{1:N})=\frac{1}{N}\sum_{j=1}^{N}u^{j}$, which clearly satisfies \eqref{eq:lambda}.}

Next, we introduce a specific concept of approximate equilibrium for $\mathcal{P}_{N}$, where followers play a Nash game among themselves, given an announced strategy of the leader. We use the notation $b^{-i}$ to denote all among $b^1, \ldots, b^N$, save $b^i$ for any variable of interest $b$. 

\begin{definition}[{$\epsilon$-Stackelberg Equilibrium}]
\label{eq:gof}
Given $\epsilon=(\epsilon^0, \hat{\epsilon}) \geq 0$, a strategy profile of admissible control strategies $\gamma^{0\star:N\star}$ for $\mathcal{P}_N$ with leader's information structure in $\{I^0_\CS, I^0_\PCS, I^0_\NCS\}$ constitutes an $\epsilon$-Stackelberg equilibrium, if
\begin{align}
&J_{N}^{0}({\gamma}^{0\star},\gamma^{1\star:N\star} )\leq \inf_{{{\gamma}^{0}}\in {{\Gamma}^{0}}}\inf_{\gamma^{1:N} \in R^{\hat{\epsilon}}(\gamma^{0})}
J_{N}^{0}({\gamma}^{0},\gamma^{1:N}) + \epsilon^{0},
\end{align} 
where $\Gamma^0 \in \{\Gamma^0_\CS, \Gamma^0_\PCS, \Gamma^0_\NCS\}$ is the corresponding set of admissible strategies for the leader, and 
$R^{\hat{\epsilon}}(\gamma^{0})$ is given by
\begin{align*}
&R^{\hat{\epsilon}}(\gamma^{0}):=\left\{\hat{\gamma}^{1:N}\in \prod_{i=1}^{N} \Gamma^{i} \middle\vert J^{i}_{N}(\gamma^{0}, \hat{\gamma}^{1:N})\leq \inf_{\gamma^{i}\in \Gamma^{i}} J^{i}_{N}(\gamma^{0}, \hat{\gamma}^{-i}, \gamma^{i})+\hat{\epsilon} \quad \forall i=1, \ldots, N \right\}.
\end{align*}
\end{definition}

This equilibrium concept is hierarchical. Given a leader's strategy, we require an approximate Nash equilibrium in the game among the $N$ followers. Then, the leader chooses a strategy that approximately optimizes her cost, accounting for the best possible (in the optimistic sense) approximate Nash response from the followers, i.e., those strategies of the followers in $R^{\hat{\epsilon}}$ for which the leader incurs the lowest cost. 

\begin{remark}
The pessimistic counterpart to the above definition has also been considered in the literature with supremum replacing infimum over $R^{\hat{\epsilon}}$ in the above definition (see, e.g., \cite{basols99}), that is, the leader selects a strategy that approximately optimizes her cost, accounting for the worst possible approximate Nash response from the followers. If the exact ($\epsilon = 0$) Nash best response set $R^{\epsilon}$ of the followers for each admissible strategy of the leader is a singleton, then the pessimistic and optimistic definitions coincide. However, the uniqueness of response sets requires restrictive assumptions, especially in the context of approximate Nash best responses. In the sequel, we consider the optimistic equilibrium concept but suppress the qualifier ``optimistic'' for brevity.
\end{remark}

%
%

Next, we introduce a notion of optimality for a strategy profile from the leader's vantage point.
\begin{definition}[{$\epsilon_{0}$-{Leader-Optimality}}]\label{def:Team} Given $\epsilon_{0}\geq0$, a strategy profile of admissible strategies $\gamma^{0\star:N\star}$ for $\mathcal{P}_{N}$ with leader's information structure $\{I^0_\CS, I^0_\PCS, I^0_\NCS\}$ constitutes an $\epsilon_{0}$-leader-optimal solution, if
\begin{equation}\label{eq:sip}
J_{N}^{0}(\gamma^{0\star:N\star})\leq \inf_{\gamma^{0:N} \in \prod_{i=0}^{N}\Gamma^{i} }J_{N}^{0}(\gamma^{0:N}) + \epsilon_{0},
\end{equation}
where $\Gamma^0$ is $\Gamma^0_\CS$, $\Gamma^0_\PCS$, or $\Gamma^0_\NCS$.
\end{definition}

In the definition of leader-optimality, we do not require the followers' responses to constitute a Nash response as we required in the Stackelberg equilibrium definition. Also, the definition does not depend on the followers' costs either. The right-hand side of \eqref{eq:sip} rather defines the optimal cost of a decentralized control problem where all players seek to minimize the leader's costs within their respective admissible policy sets. We additionally note that the leader's optimal performance under any of the leader's information structures in $\{I^0_\CS, I^0_\PCS, I^0_\NCS\}$ is identical since any leader's strategy that utilizes the controls of followers can be represented as an explicit function of observations alone, attaining the same performance.

Recall that our approach to Stackelberg equilibrium characterization is indirect; we construct a strategy, starting from an (approximate) leader-optimal strategy for the leader. Our interest lies in finding strategies for the leader that induce a desired behavior among the followers when they play according to (approximate) Nash equilibrium, which together with the leader's strategy, becomes approximately team-optimal for the leader.  In other words, the leader exploits her dynamic information structure to provide an ``incentive'' for the followers to induce a desired behavior among them, attaining her (approximate) team-optimal cost, given by the left-hand-side of \eqref{eq:sip}. We formalize such strategies in the next definition. 

\begin{definition}[{$\epsilon$-Incentive Strategy}]
Given $\epsilon=(\epsilon_{0}, \hat{\epsilon})\geq 0$, a strategy $\gamma^{0\star}$ of the leader for $\mathcal{P}_N$ with leader's information structure of $I^0_\CS$ or $I^0_\PCS$ is $\epsilon$-incentive, if there exist strategies $\gamma^{1\star:N\star}$ of followers such that $\gamma^{0\star:N\star}$ are $\epsilon_{0}$-leader-optimal and constitute an $\epsilon$-Stackelberg equilibrium.
\end{definition}
We refer to the incentive strategy $\gamma^{0\star}$ together with strategies $\gamma^{1\star:N\star}$ of the followers as $\epsilon$-{\it Stackelberg-incentive equilibrium}. {We note that although the leader's optimal performance under any of the leader's information structures in $\{I^0_\CS, I^0_\PCS, I^0_\NCS\}$ is identical, the leader's performance under incentive strategies with the leader's information structures of $\{I^0_\CS, I^0_\PCS, I^0_\NCS\}$ might not be the same. This is because the leader's optimal performance might not be realizable via a Stackelberg equilibrium under $\{I^0_\PCS, I^0_\NCS\}$ although as we will show in Theorem \ref{the:sym2}, a Stackelberg equilibrium exists that attains the leader's optimal performance under $I^0_\CS$.}

In Section \ref{sec:sym}, similar to the analysis with just one follower in \cite{bacsar1984affine}, we show that such an incentive strategy exists for $\mathcal{P}_{N}$. As it has been emphasized in the introduction, one of our goals is to study $\mathcal{P}_{N}$ when $N$ goes to infinity (the mean-field limit). 
In Section \ref{sec:main}, we show that the equilibrium strategies for $\mathcal{P}_N$ can fail to converge when $N \to \infty$, even for simple games. In Section \ref{sec:limitthe}, we modify the structure of the game and establish convergence results. In Section \ref{exists}, we provide sufficient conditions for the existence of an approximate (randomized) mean-field incentive strategy, and in Section \ref{exists}, provide an approximation result for the corresponding finite population game. Appendices include proofs of the main results. In the next section, we characterize some crucial properties of incentive strategies, including symmetry and continuity that play a central role in our analysis of the games in the infinite-population limit.

\section{Characterization of Incentive Strategies for  Stochastic Symmetric Stackelberg Games.}\label{sec:sym}

As indicated earlier, our main interest lies in studying $\mathcal{P}_{N}$ with a large number of followers. In mean-field Nash games, it is common practice to focus on \emph{symmetric} equilibria and analyze the same in the infinite population limit, e.g., see \cite{carmona2018probabilistic}. Recognizing the importance of symmetry in such analysis, we begin our study of $\mathcal{P}_N$ by addressing the question: \emph{when do symmetric incentive strategies exist?}
To that end, we need to answer two questions. The first one is whether the leader can restrict her search to {\it permutation-invariant} strategies, and yet incur no or negligible loss in team (leader) optimality. By permutation-invariant, we mean those leader strategies that do not discriminate between followers.
Without imposing any constraint on the cost function, there may be advantages for the leader to discriminate. We identify conditions in Theorem \ref{the:symmetric}, under which discrimination is not advantageous for the leader. The second question is whether such permutation-invariant leader strategies lead to a symmetric approximate Nash best response among the followers. Even if the leader announces a permutation-invariant strategy, there may not exist a symmetric Nash response for the followers. In general, there is no guarantee for symmetric games to admit symmetric Nash equilibrium (see e.g., \cite{fey2012symmetric, cheng2004notes, becker2006existence}). In the literature on mean-field Nash games, one can circumvent this problem by considering mixed strategies, in which a symmetric equilibrium exists under mild conditions (see e.g., \cite[Theorem 8.4]{cardaliaguet2010notes}). {We, however, need the existence of equilibrium in pure strategies. With mixed strategies, the leader has to know the independent randomization mechanism of followers to be able to compute her incentive strategy--a tall task in our setup. Consequently, classical fixed-point theorem-based arguments that are often used to establish the existence of equilibrium within mixed strategies in mean-field Nash games \cite{saldi2018markov, saldi2020approximate} cannot be carried over to address the question we have raised.}\footnote{In Section \ref{sec:limitthe}, we introduce a game with a major follower, for which we consider mixed strategies of followers in the mean-field limit in Section \ref{exists}. With a major player, allowing mixed strategies does not require the leader to know the randomization scheme, contrary to the case without a major player.} In Theorem \ref{the:symmetric}(ii), we provide conditions under which the second question can be answered in the affirmative.

To present our result, we need additional notation. A leader strategy $\gamma^0$ is said to be permutation invariant, if 
\begin{flalign}
&\gamma^{0}(y^{0}, y^{\sigma(1)}, \dots, y^{\sigma(N)}, u^{\sigma(1)},\dots u^{\sigma(N)})=\gamma^{0}(y^{0}, y^{1}, \dots, y^{N}, u^{1},\dots u^{N})\label{eq:PS}
\end{flalign} 
for every $y^{0}$ and any permutation $\sigma$ of the set $\{1, \dots, N\}$. Let us denote the set of all such admissible strategies by $\Gamma^{0, \PI}$.
Define $\Gamma^{\SYM}$ as the set of all symmetric admissible strategies $(\gamma, \ldots, \gamma)$ of the followers. We call a strategy profile  $\gamma^{0:N}$ symmetric if $\gamma^{0}\in \Gamma^{0, \PI}$ and $\gamma^{1:N}\in \Gamma^{\SYM}$. We are now in a position to present our first result (with proof in Appendix \ref{sec:B}) to respond to the questions raised above; the assumptions needed for that result are listed below.

\begin{assumption}\label{eq:syme}
\hfill
\begin{enumerate}[label=(\roman*)]
\item  $\mathbb{U}^{0}$ is convex, and $\mathbb{U}$ is convex and compact.
\item  $\tilde{c}^{0}(\omega_{0}, u^{0:N}):=c^{0}(\omega_{0}, u^{0}, \Lambda(u^{1:N}))$ is jointly convex in $u^{0:N}$ for every $\omega_{0}$. 
\item  $y^{1:N}$ are exchangeable, conditioned on $y^{0}, \omega_{0}$\footnote{For every $ y^{0}, \omega_{0}$ and for any permutation $\sigma$ of $\{1, \dots, N\}$, $\mathcal{L}(y^{1}, \ldots, y^{N}\vert y^{0},\omega_{0})=\mathcal{L}(y^{\sigma(1)}, \ldots, y^{\sigma(N)}\vert y^{0}, \omega_{0})$ , where $\mathcal{L}$ denotes the law of the random variables.}.
\end{enumerate}
\end{assumption}

\begin{assumption}\label{eq:symepart2}
\hfill
\begin{itemize}
\item [(i)] $c(\omega_{0}, \cdot, \cdot, \cdot)$ is continuous for every $\omega_{0}$, and $\Lambda$ is  continuous in all its arguments.
\item [(ii)] $y^{1:N}$ are i.i.d., conditioned on $y^{0}$ and $\omega_{0}$. Let $\nu$ be the conditional distribution of  $y^{i}$ on $\omega_{0}$ and $y^{0}$ for every $i\in \mathcal{N}$. For every $i=1,\ldots,N$, there exists an atomless\footnote{That is, for every Borel set $B$ with $\tilde{\nu}(B)>0$, there is another Borel set $C\subset B$ such that $\tilde{\nu}(C)>0$.} probability measure $\tilde{\nu}\in \mathcal{P}(\mathbb{Y})$, and a measurable function $h$ such that for any Borel set $A$ on $\mathbb{Y}$,
\begin{flalign*}
\nu(A\vert y^{0}, \omega_{0})&=\int_{A} h(y^{i}, y^{0}, \omega_{0})\tilde{\nu}(dy^{i}).
\end{flalign*}
\item [(iii)] $c^{0}$ and $c$ are uniformly bounded.
\end{itemize}
\end{assumption}
 


\begin{theorem}\label{the:symmetric}
Consider $\mathcal{P}_{N}$ under the leader's information structure $I_{\CS}^{0}$ or $I_{\PCS}^{0}$. 
\begin{itemize}
\item [(i)] Under Assumption \ref{eq:syme}, for every $\epsilon_{0}\geq0$,  the set of $\epsilon_{0}$-leader-optimal strategy profiles  contains a symmetric profile $({\gamma}^{0\star}, {\gamma}^{\star}, \dots, {\gamma}^{\star})\in \Gamma^{0,\PI} \times \Gamma^{\SYM}$.

\item [(ii)] If Assumptions \ref{eq:syme} and  \ref{eq:symepart2} hold, 
 then for every $\epsilon_{0}>0$, there exists a symmetric $\epsilon_{0}$-leader-optimal strategy profile  $(\tilde\gamma^{0\star}, \tilde\gamma^{\star}, \dots, \tilde\gamma^{\star})$, with $\tilde\gamma^{0\star}$ continuous in followers' actions, such that a symmetric pure $\hat{\epsilon}$-Nash best response strategy of followers exists for $\tilde\gamma^{0\star}$, i.e., $R^{\hat{\epsilon}}(\tilde\gamma^{0\star})\cap\Gamma^{\SYM}\not= \emptyset$. 

\end{itemize} 
\end{theorem}

Part (i) of Theorem \ref{the:symmetric} implies that to search for (approximate) leader-optimal strategies, one can restrict the search to symmetric strategies without any loss for the leader. In our proof, we start with a possibly asymmetric leader-optimal strategy profile, and then, we leverage exchangeability of followers and the convexity of the leader's cost function to construct a symmetric strategy profile that performs no worse for the leader.

Recall that our interest lies in finding incentive strategies for the leader, that are leader-optimal and induce a Stackelberg equilibrium. Theorem \ref{the:symmetric}(i) only addresses symmetry of a leader-optimal solution; part (ii) implies the existence of symmetric approximate pure Nash best response strategies among the followers for permutation-invariant strategies of the leader. In our proof, we first utilize Lusin's theorem and Tietze's extension theorem (see \cite[Theorem 7.5.2]{Dud02} and \cite[Theorem 4.1]{Tietze}) to show that leader's strategies can be assumed to be continuous in the actions of the followers under the dynamic information structure. Then, for such continuous symmetric strategies of the leader, we show that there exists a symmetric mixed Nash best response of the followers, where followers can independently randomize among their strategies. Finally, we use a denseness argument under a suitable topology, similar to the one used in \cite{milgrom1985distributional}, to establish that there exists an approximate pure Nash best response for the followers.

While Theorem \ref{the:symmetric}(ii) provides the existence of symmetric approximate Nash responses of the followers, an additional challenge remains to extend the result to establish the existence of symmetric Stackelberg equilibria. Note that the symmetric (approximate) Nash best response strategies of the followers may not constitute the {\it best-case} Nash equilibrium (or the {\it worst-case}) from the leader's perspective. Consequently, it becomes difficult to establish that these strategies yield an incentive strategy in the optimistic (or pessimistic) sense. One way to guarantee the same is uniqueness of the Nash best response. Establishing uniqueness of the response requires stronger conditions. These intricacies of the Nash response set $R(\gamma^{0})$ do not arise when there is only one follower (i.e., $N=1$), which has been analyzed in \cite{bacsar1984affine}. Uniqueness of (optimal) best response strategies for the single follower case can be established under mild conditions (strict convexity of the follower's cost). In the following, in the absence of uniqueness, we provide sufficient conditions under which a symmetric Nash best response strategy in fact constitutes an {optimistic} Stackelberg equilibrium. 


Following Theorem \ref{the:symmetric}, there exists an approximate symmetric leader-optimal strategy for the leader with an approximate symmetric Nash best response among followers. In the following, we start with this approximate symmetric leader-optimal strategy and use an argument similar to the one used in \cite{bacsar1984affine} to arrive at an existence and characterization result for symmetric Stackelberg-incentive equilibrium strategies for $\mathcal{P}_{N}$,  a proof of which can be found in Appendix \ref{sec:C}.  We first introduce the following assumption.
\begin{assumption}\label{assump:the3.2}
\hfill
\begin{enumerate}[label=(\roman*)]
    \item [(i)] $c$ and $\Lambda$ are such that $\tilde{c}(\omega_{0}, u^{0:N}):=c\left(\omega_{0}, u^{0}, \Lambda(u^{1:N})\right)$ is jointly strictly convex in $u^{0:N}$.
    
    \item [(ii)] $c(\omega_{0}, \cdot, \cdot, \cdot)$ is continuously differentiable for every $\omega_{0}$, and $\Lambda$ is continuously differentiable in all arguments.
\end{enumerate}
\end{assumption}

\begin{theorem}\label{the:sym2}
Consider $\mathcal{P}_{N}$ with $I^0_{\CS}$ as the leader's information structure. Suppose Assumption \ref{assump:the3.2} holds. Let $\epsilon_{0}\geq 0$ and $(\gamma^{0\star}, \gamma^{\star}, \dots, \gamma^{\star})$ constitute an $\epsilon_{0}$-leader-optimal strategy profile with leader's information structure $I^0_{\NCS}$, for which
\begin{flalign}
 \mathbb{E}\left[\nabla_{u^{0}} c\left(\omega_{0},u^{0}, \gamma^{\star}(y^{i}),\Lambda\left(\gamma^{\star}(y^{1}), \dots, \gamma^{\star}(y^{N})\right)\right)\bigg\vert y^{i}\right]_{u^{0}=\gamma^{0\star}(y^{0:N})}\not= 0,\label{eq:affect}
\end{flalign}
for any $y^{i}, i=1,\ldots,N$. Then, there exists  $\tilde{\gamma}^{0\star}\in \Gamma_{\CS}^{0}$ for the leader, given by
\begin{flalign}
\tilde{\gamma}^{0\star}(y^{0:N}, u^{1:N})&= \gamma^{0\star}(y^{0:N}) + \frac{1}{N}\sum_{i=1}^{N}Q(y^{i}, y^{0}, y^{-i})\left[u^{i}-\gamma^{\star}(y^{i})\right],\label{eq:incentive3lch3}
\end{flalign}
which, together with a symmetric (pure) Nash best response strategies $(\gamma^{\star}, \dots, \gamma^{\star})$ for the followers, constitutes an ${\epsilon}$-Stackelberg-incentive equilibrium for $\mathcal{P}_{N}$ with ${\epsilon}:=(\epsilon_{0}, 0)$, for some Borel measurable function $Q$. In addition, if each strategy in $(\gamma^{0\star}, \gamma^{\star}, \dots, \gamma^{\star})$ is weakly continuous\footnote{in the weak topology, see e.g., \cite{conway2019course}}, then $Q$ is weakly continuous.
\end{theorem}
In Theorem \ref{the:sym2}, we start with an approximate symmetric leader-optimal strategy, where Theorem \ref{the:symmetric} provided sufficient conditions. Then, we use the Hahn-Banach theorem to construct a leader's strategy (affine in the followers' actions) for which the followers' symmetric Nash equilibrium response is the desired (by the leader) optimal solution. We note that our equilibrium characterization in the Stackelberg setup is indirect, since we construct an (approximate) incentive strategy from an (approximate) leader-optimal solution.

Three remarks are now in order.
First,  the (approximate) incentive strategy is affine, and hence, continuous in followers' actions. In other words, the penalty that a follower faces for deviating from $\gamma^\star$, varies continuously with the extent of the deviation $u^i - \gamma^\star(y^i)$. Such strategies are desirable, as opposed to ``threat'' strategies, where the penalty for {any deviation} from $\gamma^\star$ by any player is heavily penalized. Second, condition \eqref{eq:affect} enables the leader to be able to influence the cost of each follower via her announced strategy, at least locally. Without such a condition, one cannot hope to construct an incentive strategy from this leader-optimal solution as the leader loses the ability to penalize deviations from the desired response. Third, Theorem \ref{the:sym2} designs an incentive strategy $\tilde{\gamma}^{0\star} \in \Gamma^{0}_{\CS}$, i.e., with access to \emph{all} followers' actions. This requirement can be relaxed to one with partial control sharing in the special case that $\hat{\Lambda}(u^{1:N})=\Lambda(u^{1:N})=\frac{1}{N}\sum_{i=1}^{N}u^{i}$. Specifically, using an argument similar to that in \cite{cansever1985stochastic}, one can identify sufficient conditions under which there exists an approximate Stackelberg-incentive equilibrium for $\mathcal{P}_{N}$ under the leader's information structure $I_{\PCS}^{0}$, of the form 
\begin{align}\label{eq:9-inc}
    \widetilde{\gamma}^{0\star}(y^{0:N},\frac{1}{N}\sum_{i=1}^{N}u^{i} )= \gamma^{0\star}(y^{0:N}) + \widehat{Q}(y^{0:N})\left[\frac{1}{N}\sum_{i=1}^{N}u^{i}-\frac{1}{N}\sum_{i=1}^{N}\gamma^{\star}(y^{i})\right]
\end{align}
for some Borel measurable function $\widehat{Q}$.

\section{Ill-posedness of Incentive Strategies in the Large Population Limit.}\label{sec:main}

We now study incentive strategies for $\mathcal{P}_{N}$ when the number of followers goes to infinity. We know that under some general conditions, the limit of the sequence of equilibrium strategies of a symmetric Nash game with finitely many players is a well-defined strategy in the infinite population game and that this limit leads to existence and characterization of mean-field equilibria \cite{lacker2018convergence, carmona2018probabilistic, fischer2017connection, lacker2016general, saldi2020approximate, saldi2018markov}. In the following, we show that the same type of result \emph{does not hold} for the stochastic incentive Stackelberg game $\mathcal{P}_{N}$ by providing a counterexample of a quadratic Gaussian (QG) game. 

Let the cost functions of the leader and followers be given as
\begin{flalign}
c^{0}\left(\omega_0, u^0, \frac{1}{N}\sum_{k=1}^{N}u^{k}\right)
&:= r^{0}(u^{0})^{2}+q^{0}\left(u^{0} + \omega_{0} + \frac{1}{N}\sum_{k=1}^{N}u^{k}\right)^{2},
\label{eq:lqgcostl}
\\
c\left(\omega_0, u^0, u^{i}, \frac{1}{N}\sum_{k=1}^{N}u^{k}\right)
&:= r(u^{i})^{2}+q\left(u^{i} + u^{0}+ \omega_{0} + \frac{1}{N}\sum_{k=1}^{N}u^{k}\right)^{2}
\label{eq:lqgcostf}
\end{flalign} 
for some real numbers $r^{0}, r, q^{0}, q>0$. Let the observations of the leader and the followers be given by $y^{i}=\omega_{0}+w^{i}$ for $i=0,\dots, N$, where $\omega_{0}$, $w^{0:N}$ are independent, and they each follow the standard normal distribution. 
Let the information structure of the leader be $I_{\PCS}^{N}$ with $\Lambda(u^{1:N}):=\frac{1}{N}\sum_{i=1}^{N}{u^{i}}$. For this setting, our assumptions of the preceding section apply, except for the compactness of strategy spaces. We state without proof that the compactness requirement can be relaxed for such quadratic Gaussian games for the conclusions to hold. Also, unique Nash best responses of the followers can be established via Banach-fixed point theorem using a technique similar to that in \cite{basar1978decentralized}.

{For the above QG, we focus on the exact leader-optimal solution that exists and is unique.} The unique leader-optimal solution with only observation-sharing can be obtained by solving the following stationarity conditions $\mathbb{P}$-a.s., 
\begin{flalign*}
 \nabla_{u^{0}}\mathbb{E}\left[ c^{0}\left(\omega_{0},u^{0},\frac{1}{N}\sum_{p=1}^{N}\gamma^{\star}_{N}(y^{p})\right) \Bigg\vert~ y^{0:N}\right]_{u^{0}={\gamma}^{0\star}_{N}(y^{0:N})}= 0,\\
 \nabla_{u^{i}}\mathbb{E}\left[ c\left(\omega_{0},{\gamma}^{0\star}_{N}\left(y^{0:N}\right), u^{i},\frac{1}{N}\left[\sum_{p=1, \not=i}^{N}\gamma^{\star}_{N}(y^{p})+u^{i}\right]\right) \Bigg\vert~ y^{i}\right]_{u^{i}=\gamma^{\star}_{N}(y^{i})}= 0
\end{flalign*}
for every $i=1, 2, \dots, N$. Hence, the unique leader-optimal solution (with only observation-sharing) takes the form,
\begin{align}
    {\gamma}^{0\star}_{N}(y^{0:N}) = \alpha_{0}^{N} y^{0} + \frac{1}{N}\sum_{i=1}^{N} \alpha^{N}y^{i}, \quad \gamma^{\star}_{N}(y^{i})=\beta^{N}y^{i},
\end{align}
with the parameters 
\begin{gather}
\alpha_{0}^{N}=-\frac{q^{0}}{2(r+q^0)},
\quad
\beta^{N}=-\frac{1}{r^{0}}\left[\frac{N(1+\alpha_{0}^{N})(r^{0}+q^{0})}{N+1} - \frac{N}{N+2}q^{0}\right],\nonumber\\
\alpha^{N}=-\frac{q^{0}}{r^{0}}\left(\beta^{N}+\frac{N}{N+2}\right).
\label{eq:QG1.ab}
\end{gather}
 For the QG game, the information structure of the leader is $I^0_{\PCS}$, and hence, we construct a symmetric leader's strategy $\tilde{\gamma}^{0\star}_{N}$ of the form in \eqref{eq:9-inc} as follows:
\begin{flalign*}
\tilde{\gamma}^{0\star}_{N}\bigg(y^{0:N}, \frac{1}{N}\sum_{i=1}^{N}u^{i}\bigg)&= \alpha_{0}^{N} y^{0} + \frac{1}{N}\sum_{i=1}^{N} \alpha^{N}y^{i} + Q_{N} \bigg[\frac{1}{N}\sum_{i=1}^{N}u^{i}-\frac{1}{N}\sum_{i=1}^{N}\beta^{N}y^{i}\bigg].
\end{flalign*}
and solve for $Q_{N}$ from the stationarity condition $\mathbb{P}$-a.s., 
\begin{flalign}
 \nabla_{u^{i}}\mathbb{E}\left[ c\left(\omega_{0},\tilde{\gamma}^{0\star}_{N}\left(y^{0:N},\frac{1}{N}\sum_{i=1}^{N}u^{i}\right), u^{i},\frac{1}{N}\left[\sum_{p=1, \not=i}^{N}\gamma^{\star}_{N}(y^{p})+u^{i}\right]\right) \Bigg\vert~ y^{i}\right]= 0\label{eq:Qstationary},
\end{flalign}
at ${u^{i}=\gamma^{\star}_{N}(y^{i})}$ for every $i=1, 2, \dots, N$. This computation yields
\begin{gather}
Q_{N}=-\frac{N(r\beta^{N}+1)+1}{\frac{1}{2}(1+\alpha_{0}^{N})+\frac{N+1}{2N}\alpha^{N}+ \frac{3N+1}{2N}\beta^{N}},
\label{eq:QN.def}
\end{gather}
where we have used the facts:
\begin{align*}
\mathbb{E}[\omega_{0}\vert y^i]=\frac{1}{2}y^{i},
\quad
\mathbb{E}[\omega_{0}\vert y^{0}]=\frac{1}{2}y^{0}, 
\quad \mathbb{E}[\omega_{0}\vert y^{0:N}]=\frac{1}{N+2}\sum_{i=0}^{N}y^{i}.
\end{align*}
Then, $(\tilde{\gamma}^{0\star}_{N},\gamma^{\star}_{N}, \dots, \gamma^{\star}_{N})$ with parameters defined in \eqref{eq:QG1.ab} and \eqref{eq:QN.def},
constitutes a Stackelberg-incentive equilibrium.
We now investigate this sequence of equilibrium strategies as $N$ goes to infinity. In the limit, we have
\begin{align*}
 \alpha^{\infty}_{0}=-\frac{q^{0}}{2(r+q^0)},
 \beta^{\infty}=-\frac{1}{r^{0}} [(1+\alpha_{0}^{\infty})(r^{0}+q^{0})-q^{0}],
 \alpha^{\infty}=-\frac{q^{0}}{r^{0}}(\beta^{\infty}+1).
\end{align*}
{We observe that 
\begin{align*}
\mathbb{E}\left[\left\vert\frac{\partial}{\partial z}\tilde\gamma_{N}^{0\star}(y^{0:N}, z)\right\vert^{2}\right]_{z=\frac{1}{N}\sum_{i=1}^{N}\gamma^{\star}_{N}(y^{i})}=\left\vert  Q_{N}\right\vert^2
\end{align*}
where this quantity represents \emph{energy} of the incentive strategy. 
Following \eqref{eq:QN.def},  $Q_{N}$ diverges as $N\to \infty$, which implies that the limiting strategy does not have finite energy. In other words, the limit of the incentive strategy with $N\to\infty$ is not a well-defined policy.} This, however, does not rule out the existence of other continuous incentive strategies that might converge and admit a mean-field limit. In the next result, we show that there does not exist any sequence of \emph{differentiable} incentive strategies that converge in the mean-field limit.

\begin{proposition}\label{the:neg1}
For the QG game with $I^0_{\PCS}$ as the leader's information structure, there is no sequence of incentive strategies $\tilde\gamma_{N}^{0\star}\in \Gamma_{\CS}^{0}$, differentiable in the average of actions of followers, converges as $N\to \infty$ 
and satisfies
\begin{flalign}
\label{eq:finiteenergy}
\limsup_{N \to \infty}\ \mathbb{E}\bigg[\bigg\vert\frac{\partial}{\partial z}\tilde\gamma_{N}^{0\star}(y^{0:N}, z)\bigg\vert^{2}\bigg]_{z=\frac{1}{N}\sum_{i=1}^{N}\gamma^{\star}_{N}(y^{i})}<\infty.
\end{flalign}
\end{proposition}

Our proof, given in Appendix \ref{sec:D}, utilizes the fact that satisfaction of the stationarity condition for all followers requires incentive strategies of the leader for which the energy grows with $N$. This is not surprising, given that as the number of followers grows, the impact of the leader's strategy on the optimization of individual followers reduces, and diminishes in the limit. To maintain even a constant level of influence on each follower, the leader must expend more ``energy'' in total--the left hand side of \eqref{eq:finiteenergy}. As a result, as $N$ goes to infinity, the energy of the leader's incentive strategy grows unbounded, yielding ill-defined limits. Our proof technique relies on differentiability of the incentive strategies; we expect the same conclusion to stand even for non-differentiable threat strategies.

\newcommand{\PNtwo}{{\mathcal{P}_N^{\textrm{maj}}}}
\newcommand{\MCS}{{{{\sf MCS}}}}
\newcommand{\ma}{{{\textrm{maj}}}}
\newcommand{\Pitwo}{{\mathcal{P}_{\infty}^{\textrm{maj}}}}

\section{A Hierarchical Incentive Stackelberg Game with a Major Follower.}
\label{sec:limitthe}

The incentive Stackelberg game setup of $\mathcal{P}_N$ from Section \ref{sec:sg} did not prove amenable to a mean-field limit with growing $N$. The key challenge lay in limiting the ``energy'' needed for the leader to monitor and incentivize an infinite population, or a constant fraction thereof, towards the leader's goals. With a growing population, the leader must therefore abandon the hope of incentivizing all followers, but rather choose to distinguish among them and incentivize finitely many of them. In this section, we take the first step toward analyzing such a game--tackle the case where the leader incentivizes a single \emph{major} follower, who plays a Nash game with a group of $N$ \emph{minor} followers. We formulate this interaction as game $\PNtwo$.

Similar to our setup in Section \ref{sec:sg}, we identify the leader as player $0$, the major follower as player $M$, and the minor followers as players $1, \ldots, N$. In $\PNtwo$, the leader begins playing by announcing a strategy. Then, all followers (major and minor) act simultaneously. We borrow the notation for the leader's and the $N$ minor followers' actions and observation spaces from Section \ref{sec:sg}. We let $\mathbb{Y}^{M}$ and $\mathbb{U}^{M}$ be subsets of finite-dimensional Euclidean spaces, endowed respectively, with their Borel $\sigma$-fields $\mathcal{Y}^{M}$ and $\mathcal{U}^{M}$, that describe the set of observations and actions of the major follower. The information available to the major follower is $I^{M}=\{y^{M}\}$. We endow the minor follower $i$ with the information set $I^{i}:=\{y^{i}, y^{M}\}$ for $i=1,\ldots,N$, and let $\Gamma^{M}$ and $\Gamma^{1}, \ldots, \Gamma^N$ denote the sets of admissible strategies of the major and minor followers that are measurable functions from their information sets to their action spaces. We consider the following
dynamic information structures for the leader.
\begin{itemize}
\item With the major follower's control being shared $I_{\MCS}^{0}:=\{y^{0},y^{M}, u^{M}\}$.
\item With only observation-sharing $I_{\NCS}^{0}:=\{y^{0},y^{M}\}$.
\end{itemize}
Accordingly, let the set of admissible strategies of the leader be denoted by $\Gamma^{0}_{\MCS}$ and $\Gamma^{0}_\NCS$  under the two information structures. The players' cost functions  are given by
\begin{flalign}
J_{N}^{0}(\gamma^{0}, \gamma^{M}, \gamma^{1:N}) &= \mathbb{E}^{\gamma^{0}, \gamma^{M}, \gamma^{1:N}}\left[c^{0}\left(\omega_{0},{u}^{0}, u^{M}, \frac{1}{N}\sum_{p=1}^{N}{u^{p}}\right)\right], 
\label{eq:1.1limitl3l}
\\
J_{N}^{M}(\gamma^{0}, \gamma^{M}, \gamma^{1:N}) &= \mathbb{E}^{\gamma^{0}, \gamma^{M}, \gamma^{1:N}}\left[c^{M}\left(\omega_{0},{u}^{0}, u^{M}, \frac{1}{N}\sum_{p=1}^{N}{u^{p}}\right)\right], 
\label{eq:1.1limitM3l}
\\
J_{N}^{i}(\gamma^{M}, \gamma^{1:N}) &= \mathbb{E}^{\gamma^{M}, \gamma^{1:N}}\left[c\left(\omega_{0},{u}^{i}, u^{M}, \frac{1}{N}\sum_{p=1}^{N}{u^{p}}\right)\right], \quad i=1,\dots, N
\label{eq:1.2limit3l},
\end{flalign}
for some Borel measurable $c^{0}: \Omega_{0} \times \mathbb{U}^{0} \times \mathbb{U}^{M} \times  \mathbb{U} \to \mathbb{R}_{+}$, $c^{M}: \Omega_{0} \times \mathbb{U}^{0} \times \mathbb{U}^{M} \times  \mathbb{U} \to \mathbb{R}_{+}$, and $c: \Omega_{0} \times \mathbb{U} \times \mathbb{U}^{M} \times  \mathbb{U} \to \mathbb{R}_{+}$. Again, $\omega_{0}$ is a cost-relevant exogenous random variable. 

In contrast to $\mathcal{P}_{N}$, the leader in $\PNtwo$ does not directly influence the costs of the minor followers through her strategy. The leader impacts the major follower's cost, who in turn, influences the minor followers through her strategy.
While the major follower plays a role that is distinct from other followers in $\PNtwo$, she is quite different from the leader in that she only wields influence over the minor followers through a Nash game, instead of through an incentive strategy that has the ability to react to the actions of these minor players. We now define an equilibrium concept for $\PNtwo$ along the lines of that for $\mathcal{P}_N$.

\begin{definition}[{$\epsilon$-Stackelberg Equilibrium}]
Given $\epsilon=(\epsilon^0, \hat{\epsilon}) \geq 0$, a strategy profile of admissible control strategies $(\gamma^{0\star}, \gamma^{M\star}, \gamma^{1\star:N\star})$ for $\PNtwo$ with leader's information structure in $\{I^0_\MCS, I^0_\NCS\}$ constitutes an  $\epsilon$-Stackelberg equilibrium for the given information structure, if
\begin{flalign}
& J_{N}^{0}(\gamma^{0\star}, \gamma^{M\star}, \gamma^{1\star:N\star}) \leq  \inf_{\gamma^{0}\in \Gamma^{0}}\inf_{(\gamma^{M}, \gamma^{1:N})\in R^{\hat{\epsilon}}(\gamma^{0})} J_{N}^{0}(\gamma^{0}, \gamma^{M}, \gamma^{1:N})+\epsilon^{0}\label{eq:stlevel2},
\end{flalign}
where $\Gamma^0 \in \{\Gamma^0_\MCS, \Gamma^0_\NCS\}$ is the corresponding set of admissible strategies for the leader and 
$R^{\hat{\epsilon}}(\gamma^{0\star})$ is the set of all strategies $(\gamma^{M\star}, \gamma^{1\star:N\star})\in \Gamma^{M}\times \prod_{i=1}^{N}\Gamma^{i}$ that satisfy \begin{flalign}
& J_{N}^{M}(\gamma^{0\star}, \gamma^{M\star}, \gamma^{1\star:N\star})\leq \inf_{\gamma^{M}\in \Gamma^{M}}J_{N}^{M}(\gamma^{0\star}, \gamma^{M}, \gamma^{1\star:N\star})+{\hat{\epsilon}}\label{eq:M1},\\
& J^{i}_{N}(\gamma^{M\star}, \gamma^{1\star:N\star}) \leq \inf_{\gamma^{i}\in \Gamma^{i}}J^{i}_{N}\left(\gamma^{M\star},(\gamma^{-i\star}, \gamma^{i})\right)+{\hat{\epsilon}}, \quad \text{for } i=1, \ldots, N.\label{eq:M2}
\end{flalign}
\end{definition}
In the above definition, we again considered the optimistic case.
Our study of  $\PNtwo$ is motivated by our desire to find leader's strategies that induce desired behavior from the followers. Lacking direct influence over minor followers, the leader in $\PNtwo$ must settle to aim for a strategy that induces a desired response from the major follower, for which the minor followers' responses are aligned with the leader's interest to the extent possible. Next, we introduce a notion of optimality for a strategy profile to capture the performance goal of the leader in $\PNtwo$.
\begin{definition}[{$\epsilon$-leader-major optimality}]
Given $\epsilon=(\epsilon^0, \hat{\epsilon}) \geq 0$, a strategy profile $(\gamma^{0\star}, \gamma^{M\star}, \gamma^{1\star:N\star})$ for $\PNtwo$ with leader's information structure of $I^0_\MCS$ or $I^0_\NCS$ constitutes an $\epsilon$-leader-major optimal solution for the given information structure, if 
\begin{flalign}
J_{N}^{0}(\gamma^{0\star}, \gamma^{M\star}, \gamma^{1\star:N\star})\leq \inf_{\gamma^{0}, \gamma^{M}\in \Gamma^{0}\times \Gamma^{M}}  \inf_{\gamma^{1:N}\in R^{\hat{\epsilon}}_{\ma}(\gamma^{M})} J_{N}^{0}(\gamma^{0}, \gamma^{M}, \gamma^{1:N})+\epsilon^{0}\label{eq:lmop},
\end{flalign}
where $\Gamma^0$ is $\Gamma^0_\MCS$ or $\Gamma^0_\NCS$ and $\gamma^{1\star:N\star}\in R^{\hat{\epsilon}}_{\ma}(\gamma^{M\star})$ are those that satisfy \eqref{eq:M2}.
\end{definition}

{In the setting of $\PNtwo$, the leader cannot attain the performance of leader-optimality for $\mathcal{P}_N$. As a result, the leader's optimal strategy involves a Nash response of the minor followers in this case. We further note that since the leader's strategy is not directly a function of the minor followers' actions and also, the minor followers' costs do not depend directly on the leader's actions, the leader's strategy in $\Gamma^0_\MCS$ admits an equivalent representation in $\Gamma^0_\NCS$, for which the response of the minor followers remains unaltered. As a result, the leader's performance under leader-major optimal strategies for $I^0_\MCS$ or $I^0_\NCS$ remains the same. As stated before, the difference arises due to the lack of leader's direct control over minor followers' actions, except through Nash response to the major follower's strategies.} To contrast the two notions of optimality, consider the following nonnegative performance loss of $\PNtwo$, compared to the leader-optimal performance.
\begin{flalign}
\mathcal{E}_{N}&=\inf_{(\gamma^{0}, \gamma^{M})\in \Gamma_{\MCS}^{0}\times \Gamma^{M}} \inf_{\gamma^{1:N} \in R_{\ma}(\gamma^{M})} J^{0}_{N}(\gamma^{0}, \gamma^{M}, \gamma^{1:N})\nonumber\\
&-\inf_{(\gamma^{0}, \gamma^{M}, \gamma^{1:N})\in \Gamma_{\MCS}^{0}\times \Gamma^{M}\times \prod_{i=1}^{N}\Gamma^{i}} J^{0}_{N}(\gamma^{0}, \gamma^{M}, \gamma^{1:N}).\label{Loss}
\end{flalign}
The first term is the best performance that the leader can expect, given that she cannot directly control the minor followers, while the second term equals the performance if she can.
We will study the behavior of this loss through an example in the sequel. 

Next, we define incentive strategies adopted to $\PNtwo$.
\begin{definition}[{$\epsilon$-Incentive Strategy}]\label{def:incentive2l}
Given $\epsilon=(\epsilon_{0}, \hat{\epsilon})\geq 0$, a strategy $\gamma^{0\star}$ of the leader for $\PNtwo$ with leader's information structure of $I^0_\MCS$ is $\epsilon$-incentive, if there exist strategies $(\gamma^{M\star}, \gamma^{1\star:N\star})$ of followers such that a profile strategy $(\gamma^{0\star}, \gamma^{M\star}, \gamma^{1\star:N\star})$ is $\epsilon_{0}$-leader-major optimal and constitutes an $\epsilon$-Stackelberg equilibrium. {We again refer to the incentive strategy $\gamma^{0\star}$ together with strategies $\gamma^{M\star}, \gamma^{1\star:N\star}$ of the followers as $\epsilon$-{\it Stackelberg-incentive equilibrium}.}.
\end{definition}

With this setup, we now address three questions. First, we ask whether the leader is able to incentivize the major follower alone in a way that the latter, together with a finite group of minor followers, play in a way that attains the performance of a leader-major optimal strategy. In other words, we aim to affirm the existence of incentive strategies for $\PNtwo$, defined above. {We again use the indirect approach which is valid for this setting as we show later on that the performance of a leader-major optimal strategy is attainable via an incentive strategy, using the facts that the leader observes the action and observation of the major follower and the costs of the minor followers do not directly depend on leader's strategy.} Second, we ask if such strategies/equilibria remain well-defined in the limit as $N \to \infty$, i.e., with a large number of followers. Third, we ask whether examples exist for which $\mathcal{E}_N = 0$, i.e., conditions on the game structure under which incentivizing the major player is enough to induce a desired response from the group of minor followers, without wielding any direct influence on them. We study the first two questions for a QG $\PNtwo$ game example in Section \ref{sec:example}. Then, in Section \ref{sec:2LMR}, we address the first two questions for general $\PNtwo$ games. Towards the end of this section, we explore the third question through an example. {Examples featuring quadratic cost structures and Gaussian observations are included to explicitly illustrate several properties outlined in the paper for general game structures.} 




\subsection{A QG Game Example of $\PNtwo$.}
\label{sec:example}

Consider a QG Stackelberg game where observations of each player are given by $y^{0}=\omega_{0}+w^{0}$
$y^{M}=\omega_{0}+w^{M}$, and $y^{i}=\omega_{0}+w^{i}$,  where $\omega_{0}$, $w^{0:N}, w^{M}$ are independent, and they each follow the standard normal distribution. We consider the leader's information structure to be $I_{\MCS}^{0}$.
Let the cost functions of the players be defined as
\begin{flalign}
&c^{0}\left(\omega_{0}, u^{0}, u^{M}, \frac{1}{N}\sum_{p=1}^{N}u^{p}\right)\nonumber\\
&\qquad \qquad :=r^{0}(u^{0})^{2}+q^{0}\left(u^{0}+u^{M}+\frac{1}{N}\sum_{p=1}^{N}u^{p}+\omega_{0}\right)^{2}+\hat{q}^{0}(u^{0}+u^{M})^{2},
\label{maj-costQG-1}\\
&c^{M}\left(\omega_{0}, u^{0}, u^{M}, \frac{1}{N}\sum_{p=1}^{N}u^{p}\right):=r^{M}(u^{M})^{2}+q^{M}\left(u^{0}+u^{M}+\frac{1}{N}\sum_{p=1}^{N}u^{p}+\omega_{0}\right)^{2}\label{maj-costQG},
\\
&c\left(\omega_{0}, u^{i}, u^{M}, \frac{1}{N}\sum_{p=1}^{N}u^{p}\right):=r(u^{i})^{2}+q\left(u^{i}+u^{M}+\frac{1}{N}\sum_{p=1}^{N}u^{p}+\omega_{0}\right)^{2}\label{maj-costQG-2},
\end{flalign}
for all $i=1, \ldots, N$, where $r^{0},r^{M},r, q^{M}>0$ and $q^{0},q, \hat{q}^{0}\geq0$. For this QG game, the unique leader-major optimal policy under the leader's information structure $I^0_\NCS$ is linear in observations.  
Consider players' strategies of the form 
\begin{flalign}\label{LQG1}
\tilde{\gamma}^{0\star}_{N}(y^{0},y^{M}, u^{M})&= \theta_{N} y^{0} + \theta^{M}_{N} y^{M} + Q_{N}[u^{M} -  \beta_{N} y^{M}],
\\
\gamma^{M\star}_{N}(y^{M})
&=\beta_{N} y^{M},
\\
\gamma^{\star}_{N}(y^{i},y^{M})&=\alpha_{N}y^{i}+\alpha^{M}_{N} y^{M} \quad \text{for } i = 1, \ldots, N.
\end{flalign}
For any arbitrary strategy $\gamma^{M}_{N}\in \Gamma^{M}$, we  characterize $R_{\ma}(\gamma^{M}_{N})$. To this end, we only need to find strategies that satisfy the stationarity criterion for each $i=1, \dots, N$, given by
\begin{flalign}
&\left(r+q+\frac{q}{N}\right)(\alpha_{N} y^{i}+\alpha^{M}_{N} y^{M})\nonumber\\
&+q \left(1+\frac{1}{N}\right)\mathbb{E}\left[\beta_{N} y^{M} +\frac{1}{N}\sum_{p=1,p\not=i}^{N}(\alpha_{N} y^{p}+\alpha^{M}_{N} y^{M})+ \omega_{0} \Bigg\vert y^{i},y^{M} \right]=0,
\label{Eq:st}
\end{flalign}
that must hold $\mathbb{P}$-a.s.
For convenience, define $r' := rN/(N+1)$. Then, \eqref{Eq:st} yields
\begin{flalign}
\alpha_{N} &= \frac{-q}{3r' + 2q\left(\frac{2N+1}{N}\right)}, 
\quad 
\alpha^{M}_{N} = -\frac{q}{r'+2q}\left(\beta_{N} + \frac{1}{3}\left( 1 + \frac{N-1}{N}\alpha_{N}\right)\right)
\label{eq:LQG3},
\end{flalign}
 where we have used the fact that $\mathbb{E}[\omega_{0}\vert y^{i}, y^{M}]=\frac{1}{3}y^{i}+\frac{1}{3}y^{M}$. We can rewrite the strategies of any minor follower $p$ in $R_{\ma}(\gamma^{M}_{N})$ as a function of $u^{M}$ and $y^{M}, y^{i}$. Again considering stationarity conditions for each minor follower, we get  
 \begin{align}\label{eq:res-min-fol}
     u^{p}=-{qu^{M}}/({r'+2q}) + F(y^{i}, y^{M})
 \end{align}
 for some function $F$, which characterizes the unique Nash response of minor followers. Hence, we replace $u^{p}$ in $c^{0}$ as above. Then, we set the expectations of the gradients of $c^{0}$ with respect to $u^{0}$ conditioned on $y^{0}, y^{M}$ and with respect to $u^{M}$ conditioned on $y^{M}$ equal to zero. For convenience, define $D:=1-q/(r^{\prime}+2q)$. Then, $\theta_{N}, \theta_{N}^{M}$, and $\beta_{N}$, given below, satisfy the preceding stationarity conditions almost surely.
 \begin{subequations}
 \begin{flalign}
\theta_{N} 
&= -\frac{q^{0}(\alpha_{N} + 1)}{3(r^{0}+q^{0}+\hat{q}^{0})},
\label{eq:LQG4}
\\
\theta^{M}_{N} &= -\frac{1}{r^{0}+q^{0}+\hat{q}^{0}}\left[(\hat{q}^{0}+q^{0})\beta_{N}+\frac{q^{0}}{3}\alpha_{N}+q^{0}\alpha^{M}_{N}\right],
\label{eq:LQG5}
\\
\beta_{N} &= -\frac{1}{3(\hat{q}^{0}+q^{0}D^{2})}\nonumber\\
&\times \left[(\hat{q}^{0}+q^{0}D)(\theta_{N}+3\theta_{N}^{M})+q^{0}D\left(1-\frac{(D+1)(N-1)}{N}\right)\alpha+q^{0}D^{2}\right]\label{eq:LQG6},
\end{flalign}
\end{subequations}
where we used the fact that $\mathbb{E}[\omega_{0}\vert y^{M}]= \frac{1}{2}y^{M}$, and $\mathbb{E}[\omega_{0}\vert y^{0}, y^{M}]= \frac{1}{3}y^{0}+ \frac{1}{3}y^{M}$.
 This implies that $({\gamma}^{0\star}_{N}, \gamma^{M\star}_{N}, \gamma^{1\star:N\star}_{N})$ with ${\gamma}^{0\star}_{N}=\theta_{N}y^{0}+\theta_{N}^{M}y^{M}$  is a leader-major optimal strategy profile, where the leader's information structure is $I_{\NCS}^{0}$. Finally, we compute $Q_{N}$ from the stationarity criterion $\mathbb{P}$-a.s.,
\begin{flalign}
\mathbb{E}\bigg[&\nabla_{u^{M}}c^{M}(\omega_{0}, \tilde\gamma^{0\star}_{N}(y^{0},y^{M}, u^{M}), u^{M}, \frac{1}{N}\sum_{i=1}^{N}\gamma^{\star}_{N}(y^{i}, y^{M}))\bigg{\vert}y^{M}\bigg]_{u^{M}=\gamma^{M\star}_{N}(y^{M})}=0, 
\label{eq:QGQ}
\end{flalign}
which yields
\begin{flalign}
-Q_{N} &= 1+ \frac{r^{M}\beta_{N}}{q^{M}L_{N}},
\quad L_{N} = (\beta_{N} + \alpha_{N}^{M} +\theta^{M}_{N}) + \frac{1}{2}(\alpha_{N} + \theta_{N} + 1) \label{eq:LQG7}.
\end{flalign}

Following \eqref{eq:res-min-fol} 
we construct $\gamma^{1\star:N\star}_{N}\in R_{\ma}(\gamma^{M\star}_{N})$, and hence, by \eqref{eq:QGQ}, we have $(\gamma^{M\star}_{N}, \gamma^{1\star:N\star}_{N})\in R(\tilde\gamma^{0\star}_{N})$, which is also unique. Then,  $\tilde{\gamma}^{0\star}_{N}$ is an incentive strategy, because
\begin{flalign}
&\inf_{\gamma^{0}\in \Gamma_{\MCS}^{0}} \inf_{(\gamma^{M}, \gamma^{1:N})\in R(\gamma^{0})} J_{N}^{0}(\gamma^{0}, \gamma^{M}, \gamma^{1:N})\nonumber\\
&\geq \inf_{\gamma^{0}, \gamma^{M}\in \Gamma_{\MCS}^{0}\times \Gamma^{M}}  \inf_{\gamma^{1:N}\in R_{\ma}(\gamma^{M})} J_{N}^{0}(\gamma^{0}, \gamma^{M}, \gamma^{1:N}),
\end{flalign} 
where the optimal performance on the right-hand side is attained by $(\tilde\gamma^{0\star}_{N}, \gamma^{M\star}_{N}, \gamma^{1\star:N\star}_{N})$. The above inequality follows from the additional infimization over $\Gamma^{M}$ on the right-hand side. 

This equilibrium sequentially converges with $N\to \infty$. Specifically, the limiting strategies are described by the following parameters with $\hat{D}:=1-q/(r+2q)$.
\begin{gather}
\begin{gathered}
\alpha_{\infty} = -\frac{q}{3r+4q},
\quad
\theta_{\infty} = -\frac{q^{0}(\alpha_{\infty} + 1)}{3(r^{0}+q^{0}+\hat{q}^{0})},
\\
\theta^{M}_{\infty} = -\frac{1}{r^{0}+q^{0}+\hat{q}^{0}}\left[(\hat{q}^{0}+q^{0})\beta_{\infty}+\frac{q^{0}}{3}\alpha_{\infty}+q^{0}\alpha^{M}_{\infty}\right],
\\
\beta_{\infty} = -\frac{\hat{q}^{0}+q^{0}\hat{D}}{3(\hat{q}^{0}+q^{0}\hat{D}^{2})}\left[\theta_{\infty}+3\theta_{\infty}^{M})\right], 
\quad
\alpha^{M}_{\infty} = -\frac{q}{r+2q}\bigg[\beta_{\infty} + \frac{1}{3}(1 + {\alpha_{\infty}})\bigg],
\\
Q_{\infty} = -1- \frac{r^{M}\beta_{\infty}}{q^{M}L_{\infty}},
\quad 
L_{\infty} = (\beta_{\infty} + \alpha_{\infty}^{M} +\theta^{M}_{\infty}) + \frac{1}{2}(\alpha_{\infty} + \theta_{\infty} + 1). 
\end{gathered}
\end{gather}
Note that $\vert Q_\infty\vert < \infty$. As a result, in the limit, the incentive strategy of the leader \emph{does} have finite energy, contrary to what we observed for $\mathcal{P}_N$. In other words, the leader can exercise a finite-energy strategy in $\PNtwo$ to incentivize the major follower, who in turn influences a large population of minor followers. This well-defined incentive strategy with $N\to \infty$ is encouraging in that  $\PNtwo$ might admit a mean-field limit. In other words, this example alludes to the possibility that the announced strategy of the leader might make a major follower play a strategy, that together with a mean-field of minor followers, would constitute a mean-field Nash equilibrium. Encouraged by our QG $\PNtwo$ game example, we formally define this game in the limit for general $\PNtwo$ instances, introduce its equilibrium concept, and study its existence in Section \ref{sec:2LMR}.

We emphasize that one must not view the incentive strategy for $\PNtwo$ as achieving the kind of leader's performance that we hoped for in $\mathcal{P}_N$. In $\PNtwo$, the leader only hopes to achieve the best she can secure by influencing the major follower, and she cannot do any better than that. We study the leader's performance loss of $\PNtwo$ compared to that of $\mathcal{P}_N$ through $\mathcal{E}_N$, defined in \eqref{Loss}. To compute $\mathcal{E}_N$, we calculate the leader-optimal solution as follows. Consider again costs defined in \eqref{maj-costQG-1}--\eqref{maj-costQG-2}. Consider the following strategies of the various players
\begin{gather}
\hat{\gamma}^{0}_{N}(y^{0},y^{M})= \hat\theta_{N} y^{0} + \hat\theta^{M}_{N} y^{M},\quad
\hat\gamma^{M}_{N}(y^{M})=\hat\beta_{N} y^{M},\nonumber\\
\hat\gamma_{N}(y^{i},y^{M})=\hat\alpha_{N}y^{i}+\hat\alpha^{M}_{N} y^{M}.
\end{gather}
Along the same line as in Section \ref{sec:main} using stationarity conditions for $c^0$ in \eqref{maj-costQG-1}, the unique leader-optimal solution $(\hat{\gamma}^{0}, \hat{\gamma}^{M}, \hat{\gamma}^{1:N})$ with only observation-sharing can be obtained by solving the following stationarity conditions $\mathbb{P}$-a.s. 
\begin{align}
\begin{aligned}
&\mathbb{E}\bigg[\nabla_{u^{0}}c^{0}(\omega_{0}, u^{0}, \hat\gamma^{M}_{N}(y^{M}), \frac{1}{N}\sum_{i=1}^{N}\hat\gamma_{N}(y^{i}, y^{M}))\bigg{\vert}y^{0}, y^{M}\bigg]_{u^{0}=\hat\gamma^{0}_{N}(y^{0},y^{M})}=0,\\
&\mathbb{E}\bigg[\nabla_{u^{M}}c^{0}(\omega_{0}, \hat\gamma^{0}_{N}(y^{0},y^{M}), u^{M}, \frac{1}{N}\sum_{i=1}^{N}\hat\gamma_{N}(y^{i}, y^{M}))\bigg{\vert}y^{M}\bigg]_{u^{M}=\hat\gamma^{M}_{N}(y^{M})}=0,\\
&\mathbb{E}\bigg[\nabla_{u^{i}}c^{0}(\omega_{0}, \hat\gamma^{0}_{N}(y^{0},y^{M}), \hat\gamma^{M}_{N}(y^{M}), \frac{1}{N}[\sum_{p=1,\not=i}^{N}\hat\gamma_{N}(y^{i}, y^{M})+u^{i}])\bigg{\vert}y^{i}, y^{M}\bigg]=0,
\end{aligned}
\end{align}
at $u^{i}=\hat\gamma_{N}(y^{i}, y^{M})$ for $i=1, \ldots, N$. Solving the above stationarity conditions, we get the relations,
\begin{subequations}
\begin{flalign}
\hat\theta_{N} &= -\frac{q^{0}}{3(r^{0}+\hat{q}^{0}+q^{0})}(1+\hat\alpha_{N}),\\
 \hat\theta^{M}_{N}&=-\frac{1}{(r^{0}+\hat{q}^{0}+q^{0})}\left[(\hat{q}^{0}+q^{0})\hat\beta_{N} + \frac{q^{0}}{3}\hat\alpha_{N}+q^{0}\hat\alpha^{M}_{N}+\frac{q^{0}}{3}\right],\\
 \hat\beta_{N} &= - \frac{1}{3}(\hat\theta_{N}+1)-\hat\theta^{M}_{N}- \hat\alpha_{N}^{M} -\frac{N-1}{3N}\hat\alpha_{N},  \\
 \hat\alpha_{N} &= - \frac{N}{N+2}[\hat\theta_{N} +1],\\
 \hat\alpha^{M}_{N} & = -\frac{\hat{q}^{0}+q^{0}}{2q^{0}}\left[\hat\theta_{N} +2\hat\theta^{M}_{N}+2\hat\beta_{N}\right] -\frac{1}{2}\hat\alpha_{N}
 \end{flalign}
\end{subequations}
that define the leader-optimal solution implicitly. {We note that the leader's optimal performance however is not attainable via an incentive strategy since $c^i$s, the costs of minor followers in \eqref{maj-costQG-2}, do not directly depend on the leader's action, and hence, \eqref{eq:affect} always fails.} Recall that the leader-major optimal performance in $\PNtwo$ matches the unique performance of Stackelberg equilibria for the QG game example, characterized in \eqref{eq:stlevel2}. To study $\mathcal{E}_N$, we choose $q^{0}=\hat{q}^{0}=q=1$ and $r^{0}=r=2$, and  depict the leader's performance under leader-major-optimal and the leader-optimal solutions as functions of $N$ in Figure \ref{Fig:1}. 
\begin{figure}[h!]
    \centering
    \includegraphics[width=0.65\textwidth]{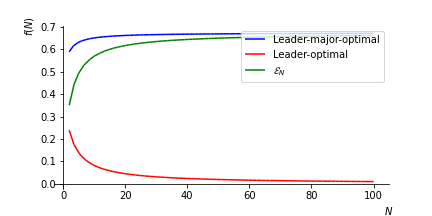}
    \caption{The leader's performance under leader-major-optimal and leader-optimal solutions.}
\end{figure}\label{Fig:1}

Per Figure \ref{Fig:1}, the leader indeed incurs an optimality loss for each $N$ in $\PNtwo$, compared to $\mathcal{P}_N$, i.e., $\mathcal{E}_N > 0$ for each $N$. As one might expect, the leader-optimal performance decreases and the leader-major-optimal performance increases as $N$ increases. This trend is not surprising, given that for leader-optimality, the increase in the number of minor followers roughly gives leader more flexibility to optimize her cost, especially in minimizing the $(u^{0}+u^{M}+\frac{1}{N}\sum_{p=1}^{N}u^{p}+\omega_{0})$ term, that helps the leader's performance. As $N\to \infty$, this term can be shown to converge to zero, allowing the leader and the major follower to apply $u^{0}=u^{M}\equiv 0$, reducing their required control efforts. On the other hand, in the leader-major optimal solution, an increase in $N$ leads to an increase in the number of players with strategic incentives that are not under the direct control of the leader, hurting the performance of the leader. In fact, an analysis similar to the leader-major optimal case allows us to conclude that the average of the minor followers' actions converges to $-\frac{1}{10} (\omega_{0}+y^{M})$, which requires a higher amount of effort from the leader and the major follower. Consequently, $\mathcal{E}_N$ defines the loss of performance that the leader expects to bear, owing to the lack of direct control over minor followers' costs.


\subsection{The Mean-Field Limit for $\PNtwo$.}\label{sec:2LMR}
We now formally introduce a Stackelberg mean-field game with one leader, one major follower, and a countably infinite number of minor followers. Adopting in this case the notation for $\PNtwo$ defined earlier, we introduce the following definitions for $\Pitwo$--the mean-field limit of $\PNtwo$. Let $R$ denote the index of a representative minor follower in the infinite population limit. Also, let $\mathcal{P}(\mathbb{Z})$ denote the space of probability measures in a space $\mathbb{Z}$, and $\mathcal{P}(\mathbb{Z}\vert\mathbb{M})$ denote the space of conditional probability measures in a space $\mathbb{Z}$ on $\mathbb{M}$.

Consider a Stackelberg game with the leader's information structure $I_{\MCS}^{0}$. Let the expected cost functions of the leader and the major/minor followers be given by
\begin{flalign}
J_{\infty}^{0}(\gamma^{0}, \gamma^{M}, \gamma^{R}) &= \mathbb{E}^{\gamma^{0}, \gamma^{M}, \gamma^{R}}\left[c^{0}(\omega_{0},{u}^{0}, u^{M}, \int_{\mathbb{U}} u d\mu)\right], \label{eq:1.1limitl3linf}\\
J_{\infty}^{M}(\gamma^{0}, \gamma^{M}, \gamma^{R}) &= \mathbb{E}^{\gamma^{0}, \gamma^{M}, \gamma^{R}}\left[c^{M}(\omega_{0},{u}^{0}, u^{M}, \int_{\mathbb{U}} u d\mu)\right], \label{eq:1.1limitM3linf}\\
J_{\infty}^{R}(\gamma^{M}, \gamma^{R}, \mu) &= \mathbb{E}^{\gamma^{M}, \gamma^{R}}\left[c(\omega_{0},{u}^{R}, u^{M}, \int_{\mathbb{U}} u d\mu)\right]\:\:\:\: \:\:\:\:\forall R\in \mathbb{N}\label{eq:1.2limit3linf},
\end{flalign}
with the condition $\mu = \mathcal{L}(\gamma^{R}(y^{R})\vert y^{M})\in \mathcal{P}(\mathbb{U}\vert\mathbb{Y}^{M})$, where $\mathcal{L}(X\vert Y)$ denotes the conditional law of a random variable $X$ on $Y$. Next, we define the corresponding equilibrium/optimality concepts for ${\Pitwo}$.

\begin{definition}[{$\epsilon$-Stackelberg-Mean-Field Equilibrium}]
Given $\epsilon=(\epsilon^0, \hat{\epsilon}) \geq 0$, a strategy profile of admissible strategies $(\gamma^{0\star}, \gamma^{M\star}, \gamma^{R\star})$ for ${\Pitwo}$ with leader's information structure in $\{I^0_\MCS, I^0_\NCS\}$ constitutes an  $\epsilon$-Stackelberg-mean-field equilibrium under the given information structure, if
\begin{flalign*}
&J_{\infty}^{0}(\gamma^{0\star}, \gamma^{M\star}, \gamma^{R\star}) \leq  \inf_{\gamma^{0}\in \Gamma^{0}} \inf_{(\gamma^{M}, \gamma^{R}, \mu)\in R^{\hat{\epsilon}, \infty}(\gamma^{0})} J_{\infty}^{0}(\gamma^{0}, \gamma^{M}, \gamma^{R})+\epsilon_{0},
\end{flalign*}
where $\Gamma^0 \in \{\Gamma^0_\MCS, \Gamma^0_\NCS\}$ is the corresponding set of admissible strategies for the leader. In the above, for every $\gamma^{0}\in \Gamma^0$,
$R^{\hat{\epsilon}, \infty}(\gamma^{0})$ is the collection of strategies $(\gamma^{M\star}, \gamma^{R\star})\in  \Gamma^{M}\times \gamma^{R}$, together with a probability measure $ \mu^{\star}\in \mathcal{P}(\mathbb{U}\vert\mathbb{Y}^{M})$, that satisfies 
\begin{flalign}
 J_{\infty}^{M}(\gamma^{0}, \gamma^{M\star}, \gamma^{R\star})&\leq \inf_{\gamma^{M}\in \Gamma^{M}}J_{\infty}^{M}(\gamma^{0}, \gamma^{M}, \gamma^{R\star})+\hat{\epsilon},\\
J^{R}_{\infty}(\gamma^{M\star},\gamma^{R\star}, \mu^{\star}) &\leq  \inf_{\gamma^{R}\in \Gamma^{R}}J^{R}_{\infty}(\gamma^{M\star}, \gamma^{R}, \mu^{\star})+\hat{\epsilon},\label{eq:2ndmft}\\
\mu^{\star} &= \mathcal{L}(\gamma^{R\star}(y^{R})\vert y^{M})\label{eq:2ndmft-consi}.
\end{flalign}
Let $R^{\hat{\epsilon}, \infty}_{\ma}(\gamma^{M\star})$  be the set of $(\gamma^{R\star},\mu^{\star})\in \Gamma^{R} \times \mathcal{P}(\mathbb{U}\vert\mathbb{Y}^{M})$ that satisfy \eqref{eq:2ndmft} and \eqref{eq:2ndmft-consi}.
\end{definition}
In the preceding definition of Stackelberg-mean-field equilibrium, the consistency condition on $\mu^{\star}$ is conditional on $y^{M}$ since minor followers have access to the common random variable $y^{M}$. In the absence of $y^{M}$, the consistency condition simplifies to $\mu^{\star}=\mathcal{L}(\gamma^{R\star}(y^{R}))$, for which our analysis remains valid.

\begin{definition}[{$\epsilon$-Leader-Major Mean-Field Optimality}]
Given $\epsilon=(\epsilon^0, \hat{\epsilon}) \geq 0$, a strategy profile $(\gamma^{0\star}, \gamma^{M\star}, \gamma^{R\star})$ for ${\Pitwo}$ with leader's information structure  $I^0_\MCS$ or $I^0_\NCS$ constitutes an $\epsilon$-leader-major mean-field optimal solution under the given information structure, if 
\begin{flalign}
J_{\infty}^{0}(\gamma^{0\star}, \gamma^{M\star}, \gamma^{R\star})\leq \inf_{\gamma^{0}, \gamma^{M}\in \Gamma^{0}\times \Gamma^{M}}   \inf_{(\gamma^{R}, \mu)\in R^{\hat{\epsilon}, \infty}_{\ma}(\gamma^{M})} J_{\infty}^{0}(\gamma^{0}, \gamma^{M}, \gamma^{R})+\epsilon^{0}\label{eq:lmopmft},
\end{flalign}
where $\Gamma^0$ is $\Gamma^0_\MCS$ or $\Gamma^0_\NCS$.
\end{definition}

{We note that since the leader's strategy is not directly a function of the minor followers' actions and also, the minor followers' costs do not depend directly on the leader's actions, a leader's strategy in $\Gamma^0_\MCS$ admits an equivalent representation in $\Gamma^0_\NCS$, for which the response of the minor followers remains unaltered. As a result, the leader's performance under leader-major mean-field optimal strategies for $I^0_\MCS$ or $I^0_\NCS$ remains the same.}

\begin{definition}[{$\epsilon$-Mean-Field Incentive Strategy}]\label{def:incentive2lmft}
Given $\epsilon=(\epsilon_{0}, \hat{\epsilon})\geq 0$, a strategy $\gamma^{0\star}$ of the leader for ${\Pitwo}$ with leader's information structure of $I^0_\MCS$ is $\epsilon$-incentive, if there exist strategies $(\gamma^{M\star}, \gamma^{R\star})$ of followers such that $(\gamma^{0\star}, \gamma^{M\star}, \gamma^{R\star})$ is $\epsilon_{0}$-leader-major mean-field optimal and constitutes an  (optimistic) $\epsilon$-Stackelberg mean-field equilibrium. We refer to the entire profile as a {\it mean-field incentive equilibrium}. 
\end{definition}

Equipped with the above definitions, we now present our main result (Theorem \ref{the:Ma-Mi} for $\PNtwo$ and $\Pitwo$), whose proof can be found in Appendix \ref{sec:E}. First, we introduce the assumptions needed for our result. 

\begin{assumption}\label{assumption:m-m3l}
\hfill
\begin{itemize}
\item [(i)] $c^{M}(\omega_{0}, \cdot, \cdot, \cdot)$ is continuously differentiable and (jointly) strictly convex for every $\omega_{0}$. 
\item [(ii)] $c(\omega_{0}, \cdot, \cdot, \cdot)$ and $c^{0}(\omega_{0}, \cdot, \cdot, \cdot)$ are continuous  for every $\omega_{0}$.
\item [(iii)] $y^{0}, y^{M}, y^{1:N}$  are independent, conditioned on $\omega_{0}$, and $y^{1:N}$  are identically distributed, conditioned on $\omega_{0}$.
\item [(iv)] $\mathbb{U}^{0}$ and $\mathbb{U}^{M}$ are compact.
\item [(v)] Let $\nu^{0}, \nu^{M}$ be respectively the conditional distributions of $y^{0}$ and $y^{M}$ on $\omega_{0}$. There exist probability measures $\tilde{\nu}^{0}\in \mathcal{P}(\mathbb{Y}^{0})$ and $\tilde{\nu}^{M}\in \mathcal{P}(\mathbb{Y}^{M})$, and measurable functions $h^{0}$ and $h^{M}$ such that for any Borel sets $A$ and $B$ on $\mathbb{Y}^{0}$ and $\mathbb{Y}^{M}$, respectively
\begin{flalign*}
\nu^{0}(A\vert\omega_{0})=\int_{A} h^{0}(y^{0},\omega_{0})\tilde{\nu}^{0}(dy^{0}),
\quad
\nu^{M}(B\vert\omega_{0})=\int_{B} h^{M}(y^{M},\omega_{0})\tilde{\nu}^{M}(dy^{M}).
\end{flalign*}
\item [(vi)] There exists $\delta_{0}, \delta_{M}, \delta>0$ such that for any admissible strategies $(\gamma^{0}, \gamma^{M}, \gamma^{1:N})$,
\begin{flalign*}
&\sup_{N\geq 1}\mathbb{E}^{\gamma^{0}, \gamma^{M}, \gamma^{1:N}}\left[\left\vert c^{0}(\omega_{0},{u}^{0}, u^{M}, \frac{1}{N}\sum_{p=1}^{N}{u^{p}})\right\vert^{1+\delta_{0}}\right]<\infty,
\\
&\sup_{N\geq 1}\mathbb{E}^{\gamma^{0}, \gamma^{M}, \gamma^{1:N}}\left[\left\vert c^{M}(\omega_{0},{u}^{0}, u^{M}, \frac{1}{N}\sum_{p=1}^{N}{u^{p}})\right\vert^{1+\delta_{M}}\right]<\infty,
\\
&\sup_{N\geq 1}\mathbb{E}^{\gamma^{M}, \gamma^{1:N}}\left[\left\vert c(\omega_{0},{u}^{i}, u^{M}, \frac{1}{N}\sum_{p=1}^{N}{u^{p}})\right\vert^{1+\delta}\right]<\infty, \quad  i = 1,\dots, N.
\end{flalign*}
\end{itemize}
\end{assumption}



\begin{theorem}\label{the:Ma-Mi}
The following assertions are true for $\PNtwo$ and $\Pitwo$.
\begin{itemize}
\item [(i)] 
Consider $\PNtwo$ with leader's information structure $I^0_{\MCS}$. 
Let Assumption \ref{assumption:m-m3l}(i) hold. Suppose that $\epsilon=(\epsilon^0, \hat{\epsilon})\geq 0$ and $(\gamma^{0\star}_{N}, \gamma^{M\star}_{N}, \gamma^{\star}_{N}, \dots, \gamma^{\star}_{N})$ constitutes an $\epsilon$-leader-major optimal strategy profile  with leader's information structure $I^0_{\NCS}$ that satisfies
\begin{flalign}\label{eq:sat}
 \mathbb{E}\bigg[\nabla_{u^{0}}c^{M}(\omega_{0},u^{0}, \gamma^{M\star}_{N}(y^{M}),\frac{1}{N}\sum_{i=1}^{N}\gamma^{\star}_{N}(y^{i}, y^{M}))\bigg\vert y^{M}\bigg]_{u^{0}=\gamma^{0\star}_{N}(y^{0}, y^{M})}\not = 0 \quad \mathbb{P}\text{-a.s.}
\end{flalign}
Then, there exists  $\tilde{\gamma}^{0\star}_{N}\in \Gamma_{\MCS}^{0}$ for the leader, given by
\begin{flalign}
\tilde{\gamma}^{0\star}_{N}(y^{0}, y^{M}, u^{M})&= \gamma^{0\star}_{N}(y^{0}, y^{M}) + Q_{N} (y^{0}, y^{M})[u^{M}-\gamma^{M\star}_{N}(y^{M})],\label{eq:incentive3l}
\end{flalign} 
which together with the followers' strategy profile $(\gamma^{M\star}_{N}, \gamma^{\star}_{N}, \dots, \gamma^{\star}_{N})$, constitutes an $\epsilon$-Stackelberg-incentive equilibrium for $\PNtwo$, for some Borel measurable function $Q_{N}(y^{0}, y^{M})$. In addition, if the strategies $\gamma^{0\star}_{N}$, $\gamma^{M\star}_{N}$, and $\gamma^{\star}_{N}$ are weakly continuous, then so is $Q_{N}(y^{0}, y^{M})$.

\item [(ii)] Let Assumption \ref{assumption:m-m3l} hold. If the sequences  $\{\tilde{\gamma}^{0\star}_{N}\}_{N}$, $\{\gamma^{M\star}_{N}\}_{N}$, and $\{\gamma^{\star}_{N}\}_{N}$ of $\epsilon$-Stackelberg-incentive equilibrium strategies for $\PNtwo$ converge point-wise as $N\to \infty$, then the limit constitutes an  $\epsilon$-mean-field incentive equilibrium for ${\Pitwo}$.
\end{itemize}
\end{theorem}

{As a result of Theorem \ref{the:Ma-Mi}, if a sequence of  (approximate) incentive strategies for $\PNtwo$ converges as $N\to\infty$, then this limit has a specific meaning. Precisely, these limiting strategies constitute an approximate equilibrium of the Stackelberg mean-field game under the leader's dynamic information structure $I_{\MCS}^{0}$. Proof of Theorem \ref{the:Ma-Mi}(i) uses an analogous argument as that used in the proof of Theorem \ref{the:sym2}.  Proof of Theorem \ref{the:Ma-Mi}(ii) follows from point-wise convergence of $\epsilon$-Stackelberg-incentive equilibrium strategies for $\PNtwo$,  convergence of the empirical measures on actions and observations of minor followers, and continuity of players' costs using the generalized dominated convergence theorem \cite[Theorem 3.5]{serfozo1982convergence}.}

    Regarding the assumption on the existence of an approximate leader-major optimal strategy profile for $\PNtwo$, we note that for $\PNtwo$ the information structures and the cost functions are asymmetric among followers in contrast to $\mathcal{P}_{N}$ due to the presence of the major follower. Also, in $\PNtwo$, the information structures of the major and the minor followers are static. Hence, under similar assumptions as those in Assumptions \ref{eq:syme} and  \ref{eq:symepart2}, using an argument utilized in the proof of Theorem \ref{the:symmetric}(ii), we can show that approximate symmetric pure Nash best response strategies for minor followers exist for all $(\gamma^0, \gamma^M)$. This guarantees the existence of an approximate leader-major optimal solution. In contrast to $\mathcal{P}_N$, we do not require any symmetry in the leader's strategy with respect to the followers--the leader does not directly interact with the minor followers through their actions and by design cannot discriminate among them.

     The set of approximate Nash best responses of the minor followers to the leader's and the major follower's strategies is always non-empty and contains symmetric strategies for the followers, owing to an argument similar to that of Theorem \ref{the:symmetric}(ii). Since the costs are bounded below by zero, this implies that an approximate leader-major optimal solution always exists. However, establishing an existence result for an (approximate) leader-major optimal solution with a guaranteed existence of an (optimistic) symmetric pure-strategy Nash response from minor followers remains challenging. This is in contrast to the setting in Section \ref{sec:sym}, where by Theorem \ref{the:symmetric}(i), the existence of an approximate leader-optimal optimal solution with symmetric strategies among followers was established. That is because, the leader-major optimal solution is still a Stackelberg equilibrium, which requires existence of an (optimistic) symmetric pure-strategy Nash response from minor followers in contrast to the leader-optimal solution, which is a globally optimal solution of the leader's team problem.  

    In Theorem \ref{the:Ma-Mi}, if in addition, for any $\gamma^{0}\in \Gamma_{\MCS}^{0}$ affine in $u^{M}$, $R(\gamma^{0})$ and $R^{\infty}(\gamma^{0})$ are singletons (e.g., QG games), then the results of Theorem \ref{the:Ma-Mi} apply to the pessimistic case as well. Akin to our discussion in Section \ref{sec:sym}, non-uniqueness of Nash best response leads to intricacies regarding whether symmetric Nash best response strategies (among minor followers) correspond to the best-case (or the worst-case) for the leader and the major follower in the leader-major optimal solution. Such difficulties can be avoided either by imposing uniqueness of Nash best response strategies among followers in  Assumption \ref{assumption:m-m3l}(vii) to affine strategies of the leader, or by establishing weaker results and focusing only on symmetric Nash response of minor followers. 

In $\Pitwo$, recall that the leader only incentivizes the major follower to induce a desired response from the infinite population of minor followers. However, such an incentive strategy typically results in performance worse than what the leader can garner from incentivizing all minor followers. In other words, $\mathcal{E}_{N} > 0$, as defined in \eqref{Loss}. In the following, we present a simple QG game example for which $\mathcal{E}_{N}=0$.  

Consider a special case of the QG game in Section \ref{sec:example} with the cost functions of the leader and minor followers given by
\begin{flalign}
c^{0}\left(\omega_{0},{u}^{0}, u^{M},  \frac{1}{N}\sum_{p=1}^{N}{u^{p}}\right)&=r^{0}(u^{0})^{2}+q^{0}\left(u^{0}+\frac{1}{N}u^{M}+\frac{1}{N}\sum_{p=1}^{N}u^{p}+\omega_{0}\right)^{2} \\
c\left(u^{i}, u^{M}, \frac{1}{N}\sum_{p=1}^{N}{u^{p}}\right) 
&= \left({u}^{i}- \frac{1}{N}\sum_{p=1}^{N}{u^{p}}\right)^{2}
+
\left({u}^{i}-u^{M}\right)^{2}\label{eq:1.2limit},
\end{flalign}
and the cost function of the major follower given by \eqref{maj-costQG}. Suppose $y^{M}=y=y^{i}$ for $i\in \{1, \dots, N\}$. First, we compute a strategy profile $(\gamma^{0\star}, \gamma^{\star}, \gamma^{\star}, \ldots, \gamma^{\star})$ that constitutes a leader-optimal solution with leader's strategy  $\gamma^{0\star}_{N}=\theta y^{0} + \theta^{M} y$ and strategies of major and minor followers as  $\gamma^{\star}(y)=\beta y$. It can be shown that $\theta, \theta^{M}, \beta$ are given uniquely by
\begin{flalign}
\theta=-\frac{q^{0}}{3(q^{0}+r^{0})}, \quad \theta^{M}=-\frac{3(\theta +1)+2q^{0}}{6(r^{0}+q^{0}-1)},\quad \beta=-\frac{\theta+1+2\theta^{M}}{2}. 
\end{flalign}
satisfies the stationarity conditions in which we set the expectation of the partial derivative of $c^0$ to zero with respect to the actions of the players conditioned on their available information. We note that this strategy is unique not only within linear strategies but also in the general class of Borel-measurable strategies since the cost is quadratic, random variables are Gaussian, and the information structure of the leader's team problem is static (i.e., it does not depend on actions of other players).
Next, we obtain an affine incentive strategy of the form given in \eqref{eq:incentive3l}, where $Q$ is computed from a stationarity condition similar to \eqref{eq:QGQ} as
\begin{flalign}
Q=-\frac{2 r^{M} \beta + q^{M}(4 \beta + 1+\theta)}{q^{M}(4\beta+1+\theta)}.
\end{flalign}
 Note that 
 \begin{align}
 \begin{aligned}
 &\nabla_{u^{i}}\mathbb{E}\bigg[c(\omega_{0}, u^{i}, \beta y, \frac{1}{N}\sum_{p=1, p\not= i}^{N}\beta y+u^{i})\bigg\vert y\bigg]_{u^{i}=\beta y}=0, \ \mathbb{P}-a.s., 
 \\
 & \implies 
  \quad \inf_{\gamma^{i}\in
\Gamma^{i}}J^{i}_{N}(\tilde{\gamma
}^{0\star}, \gamma^{\star}_{N}, \gamma^{-i\star}_{N}, \gamma^{i})=J^{i}_{N}(\tilde{\gamma
}^{0\star}_{N}, \gamma^{\star}_{N}, \dots, \gamma^{\star}_{N}).
\end{aligned}
 \end{align}
As a result, we have   $R_{\ma}(\gamma^{\star}_{N})=\{(\gamma^{\star}_{N}, \dots, \gamma^{\star}_{N})\}$, implying that the leader-major optimal solution  coincides with leader-optimal solution. Thus, $\mathcal{E}_{N}=0$ for all $N>0$, and remains so in the limit.


\section{Existence of an Approximate Randomized Mean-Field Incentive Equilibrium.}\label{exists}

Our analysis in the last section yielded the result that if a sequence of incentive equilibria for $\PNtwo$ converges as $N\to \infty$, then the limit is a mean-field incentive equilibrium for $\Pitwo$. In this section, we ask the opposite question: \emph{does a limit point of such equilibria exist as $N \to \infty$?} We answer this question in the affirmative, but when we allow all players to independently randomize their actions. In other words, we establish that mixed-strategy incentive equilibria exist for $\PNtwo$, for which an accumulation point constitutes a mean-field mixed-strategy incentive equilibrium for $\Pitwo$. 
We do so by extracting a convergent subsequence of said equilibria. Such an argument relies on compactness of strategies in a suitable topology. Function spaces are often not compact; spaces of probability measures often are, in a suitable topology. As a result, limiting arguments apply, when players employ mixed strategies, but not that easily, when players use pure strategies.
See similar arguments being employed in \cite{SSYdefinetti2020} for team problems and \cite{lacker2018convergence, fischer2017connection} for Nash games.

We allow all players--the leader, the major follower, and the collection of minor followers--to utilize independently randomized strategies. In $\PNtwo$, the leader only observes the action of the major follower. Thus, incentive strategies for the leader must account for the randomization scheme of the major follower. However, she does \emph{not} need to consider the randomization schemes of a growing population of minor followers to design her incentive strategy. Said differently, mixed strategies in $\PNtwo$ requires the leader to only track the randomization scheme of a single follower, as opposed to an unrealistic requirement to know the same for a large population of minor followers, if we allowed randomized strategies in $\mathcal{P}_{N}$.


We endow the players with the following information sets. Leader's information set is $I_{\MCS}^{0}=\{y^{0}, y^{M}, u^{M}\}$, major follower's information set is $I^{M}:=\{y^{M}\}$, and the information set of minor follower $i$ is $I^{i}:=\{y^{i}\}$ for $i=1,\ldots,N$.
Recall the following (static) information for the leader,  which is used in our analysis,  $I^0_{\NCS}:=\{y^{0}, y^{M}\}$. A randomized strategy $\pi^{0}$ for the leader under $I_{\NCS}^{0}$ is a stochastic kernel on $\mathbb{U}^{0}$ given $y^{0}, y^{M}$. A randomized strategy $\pi^{M}$ for the major follower is a stochastic kernel on $\mathbb{U}^{M}$ given $y^{M}$. Also, a randomized strategy $\pi^{i}$ for the  follower $i$ is a stochastic kernel on $\mathbb{U}$ given $y^{i}$. We denote the space of randomized strategies for the leader, the major follower, and the minor follower $i$ by $\Delta^{0}_{\NCS}$, $\Delta^{M}$, and $\Delta^{i}$, respectively. We also define sets of strategies $R^{\hat{\epsilon}}_{\pi}(\pi^0)$ and $R^{\hat{\epsilon}}_{\ma,\pi}(\pi^M)$ for mixed strategies as counterparts of those with pure strategies in \eqref{eq:M1} and \eqref{eq:M2}. We also use the notation $a \sim b$ to denote that $a$ has a distribution $b$. We introduce the following assumptions required for the ensuing analysis.
  \begin{assumption}\label{assum:5}
  \hfill
  \begin{itemize}
      \item [(i)] $c^{0}$ is uniformly bounded and $c^{0}(\omega_{0}, \cdot, \cdot, \cdot)$ is continuous for every $\omega_{0}$. 
      \item [(ii)] $c^{M}$ and $c$ are uniformly bounded, and $c^{M}(\omega_{0}, \cdot, \cdot, \cdot)$, and $c(\omega_{0}, \cdot, \cdot, \cdot)$ are continuous for every $\omega_{0}$.
      \item [(iii)] $\mathbb{U}^{M}$ and $\mathbb{U}$ are compact.
      \item [(iv)] $y^{0}, y^{M}, y^{1:N}$  are independent, conditioned on $\omega_{0}$, and $y^{1:N}$  are identically distributed, conditioned on $\omega_{0}$.
      \item [(v)] Let $\nu^{0}, \nu^{M}$, and $\nu$ be, respectively, the distributions of $y^{0}$, $y^{M}$, $y^{i}$, conditioned on $\omega_{0}$. There exist probability measures $\tilde{\nu}^{0}\in \mathcal{P}(\mathbb{Y}^{0})$, $\tilde{\nu}\in \mathcal{P}(\mathbb{Y})$, and atomless probability measure $\tilde{\nu}^{M}\in \mathcal{P}(\mathbb{Y}^{M})$, and bounded measurable functions $h^{0}$, $h$, $h^{M}$ such that for any Borel sets $A$, $B$, and $C$ on $\mathbb{Y}^{0}$, $\mathbb{Y}^{M}$, and $\mathbb{Y}$, respectively
\begin{flalign*}
&\nu^{0}(A\vert\omega_{0})=\int_{A} h^{0}(y^{0},\omega_{0})\tilde{\nu}^{0}(dy^{0}),
\quad
\nu^{M}(B\vert\omega_{0})=\int_{B} h^{M}(y^{M},\omega_{0})\tilde{\nu}^{M}(dy^{M}),\\
&\nu(C\vert\omega_{0})=\int_{C} h(y^{i},\omega_{0})\tilde{\nu}(dy^{i})~~~\forall i=1, \dots, N.
\end{flalign*}
  \end{itemize}
  \end{assumption}

In our first result with randomization allowed for players, we show that the leader's performance under any randomized leader-major optimal solution can be approximated by an approximate randomized leader-major optimal solution that enjoys pure strategies for the leader and the major follower. In other words, the leader and the major follower's strategies need not be random for approximately leader-major optimal performance, albeit allowing the minor followers to randomize. Then, we utilize this leader-optimal strategy to construct an incentive strategy through a construction that is similar to that for $\PNtwo$ without randomization. The proof of the following proposition can be found in Appendix \ref{A:F}. 
  \begin{proposition}\label{lem:exi-1}
  Consider $\PNtwo$. 
  \begin{itemize}
      \item [(i)] Let Assumption \ref{assum:5} hold. Let the leader's information structure be $I_{\NCS}^{0}$. For any $\epsilon^{0}>0$, there exists $\hat{\epsilon}>0$ such that 
  \begin{flalign}\label{eq:mainlem1}
  &\bigg\vert\inf_{\gamma^{0}, \gamma^{M}\in \Gamma^{0}_{\NCS} \times \Gamma^{M}} \inf_{\pi^{1:N}\in R^{\hat{\epsilon}}_{\ma, \pi}(\gamma^{M})} J^{0}_{N}(\gamma^{0}, \gamma^{M}, \pi^{1:N})\nonumber\\
  &\qquad \qquad \qquad  -\inf_{\pi^{0}, \pi^{M}\in \Delta^{0}_{\NCS} \times \Delta^{M}} \inf_{\pi^{1:N}\in R_{\ma, \pi}(\pi^{M})} J^{0}_{N}(\pi^{0}, \pi^{M}, \pi^{1:N})\bigg\vert\leq\epsilon^{0}.
  \end{flalign}
  \item [(ii)] With leader's information structure being $I_{\MCS}^{0}$, let Assumption \ref{assumption:m-m3l}(i) hold and there exist an $(0,\hat{\epsilon})$-leader-major optimal strategy profile $(\gamma^{0\star}, \gamma^{M\star}, \pi^{\star}, \ldots, \pi^{\star})\in \Gamma^{0}_{\MCS}\times \Gamma^{M}\times \prod_{i=1}^{N}\Delta^{i}$, that satisfies
  \begin{flalign}
  \label{eq:sat2}
\nabla_{u^{0}}\mathbb{E}\bigg[ c^{M}(\omega_{0},u^{0}, \gamma^{M\star}(y^{M}),\frac{1}{N}\sum_{i=1}^{N}u^{i\star})\bigg\vert y^{M}\bigg]_{\substack{{u^{0}=\gamma^{0\star}(y^{0}, y^{M}),}\\{u^{i\star}\sim \pi^{\star}(\cdot\vert y^{i})}}}\not = 0, \quad \mathbb{P}-a.s.
\end{flalign}
  Then, there exists a strategy profile $(\tilde{\gamma}^{0\star}, \gamma^{M\star},  \pi^{\star}, \ldots, \pi^{\star})\in \Gamma^{0}_{\MCS}\times \Gamma^{M}\times \prod_{i=1}^{N}\Delta^{i}$, symmetric among minor followers and constitutes an  $(\epsilon^{0}, \hat{\epsilon})$-Stackelberg-incentive equilibrium for $\PNtwo$ among all randomized strategy profiles $\Delta^{0}_{\MCS}\times \Delta^{M}\times \prod_{i=1}^{N}\Delta^{i}$.
  \end{itemize}
  \end{proposition}

Assumption \ref{assum:5}(iv)-(v) allows us to invoke a change of measure argument, under which the distributions of observations of the minor followers are independent. Endowing the space of joint distributions over the action and observation spaces with the $w$-$s$ topology
\footnote{The $w$-$s$ topology on the set of probability measures $\mathcal{P}(\mathbb{U}^{0}\times \mathbb{Y}^{0} \times \mathbb{U}^{M})$ is the coarsest topology under which $\int f(y^0,y^M, u^0)P(dy^0, du^0, du^M) : \mathcal{P}(\mathbb{U}^{0}\times \mathbb{Y}^{0} \times \mathbb{U}^{M}) \to \mathbb{R}$ is continuous for every measurable and bounded function $f$, which is continuous in $u^0$ for every $y^0, y^M$. Unlike the weak convergence topology, $f$ need not to be continuous in $y^0, y^M$ (see \cite{Schal}).}, 
we then  
use a denseness argument along the lines of \cite[Theorem 3]{milgrom1985distributional} to prove Proposition \ref{lem:exi-1}(i). Our proof also makes use of the property that the leader's cost is linear in her own randomized strategy $\pi^0_N$, leading to the existence of an optimum strategy that defines an extreme point of the set $\Delta^0$ that is deterministic.
The proof of part (ii) is an application of Hahn-Banach theorem as we use to deduce Theorem \ref{the:Ma-Mi}(i). 

One might surmise that the approximate incentive strategies, guaranteed per Proposition \ref{lem:exi-1}, will converge to a mean-field limit in a suitable topology, in which the strategies of the leader and the major follower are deterministic. However, proving asymptotic convergence in function spaces remains challenging. As we prove in our next result, Theorem \ref{the:Existence}(i), these deterministic incentive equilibria for $\PNtwo$ admit a subsequence that converges to an approximate incentive equilibrium for the mean-field counterpart $\Pitwo$. The existence of such an accumulation point then defines the first reason to study the game $\Pitwo$ as an approximation to $\PNtwo$ with a large number of followers. In Theorem \ref{the:Existence}(ii), we argue the converse. Regardless of how one produces it, a mean-field incentive equilibrium of $\Pitwo$, per our result, constitutes an approximate incentive equilibrium for $\PNtwo$. In other words, any incentive equilibrium of the mean-field limit provides an approximation for the behavior of the finite-population variant. 
The proof of the following result is deferred till Appendix \ref{A:H}.

\begin{theorem}\label{the:Existence}
Consider $\PNtwo$ and $\Pitwo$ with $I_{\MCS}^{0}$ as the leader's information structure. Let Assumption \ref{assum:5} hold, and ${\epsilon}=(\epsilon^{0}, \hat{\epsilon})>0$.
\begin{itemize}
    \item [(i)] Let Assumption \ref{assumption:m-m3l}(i) hold. Further suppose that for every finite $N\in \mathbb{N}$, there exists a strategy profile $(\gamma^{0\star}_{N}, \gamma^{M\star}_{N}, \pi^{\star}_{N}, \ldots, \pi^{\star}_{N})\in \Gamma^{0}_{\MCS}\times \Gamma^{M}\times \prod_{i=1}^{N}\Delta^{i}$, symmetric among minor followers, satisfying \eqref{eq:sat2}, that constitutes an $(0, \hat{\epsilon})$-leader-major optimal solution for $\PNtwo$.  Then, there exists a randomized ${\epsilon}$-mean-field incentive equilibrium for $\Pitwo$.
    \item [(ii)] Suppose that $(\pi^{0\star}, \pi^{M\star}, \pi^{\star})\in \Delta^{0}_{\MCS}\times \Delta^{M}\times \Delta^{R}$ constitutes an ${\epsilon}$-mean-field incentive equilibrium for $\Pitwo$. Then, $(\pi^{0\star}, \pi^{M\star}, \pi^{\star}, \ldots, \pi^{\star})$ constitutes an ${\epsilon}_{N}$-Stackelberg-incentive equilibrium for $\PNtwo$ among symmetric randomized strategies $\Delta_{\MCS}^{0}\times \Delta^{M}\times \Delta^{\SYM}$, where ${\epsilon}_{N}:=(\epsilon_{N}^{0}, \hat{\epsilon}_{N})>0$ such that $\epsilon_{N}^{0}\to \epsilon^{0}$ and $\hat\epsilon_{N}\to \hat\epsilon$ as $N\to \infty$.
    \end{itemize}

\end{theorem}
We shed light on the proof of Theorem \ref{the:Existence}(i). Using Proposition \ref{lem:exi-1}, there exists an approximate Stackelberg-incentive equilibrium among randomized strategies for $\PNtwo$ that admit the same with the leader and the major follower playing pure strategies, and the minor followers' randomized strategies are symmetric. We show that, under our assumptions, the sequence of empirical measures of observations and actions induced by randomized strategies of minor followers under an approximate incentive equilibrium for $\PNtwo$ is tight (pre-compact in the weak convergence topology), implying that this sequence admits an accumulation point.  Finally, we show that the set of randomized strategies for the leader and followers are compact in the $w$-$s$ topology (closed and tight).
The rest requires us to argue that this accumulation point is in fact an incentive equilibrium of the mean-field counterpart $\Pitwo$ of $\PNtwo$.

Part (ii) is a product of our analysis in part (i).  Comparing the result of part (ii) with the result of Theorem \ref{the:Ma-Mi}(i), we observe that the approximate incentive equilibrium provided in Theorem \ref{the:Existence}(ii) is independent of the number of minor followers $N$, although it might not be deterministic among the players and may not be affine in the action of the major follower for the leader. We rather start with \emph{any} randomized incentive equilibrium of the mean-field game and replicate the minor follower strategy $N$ times for \emph{any} $N$ number of such followers and argue that it defines an approximate incentive equilibrium for the finite-N variant of the game $\PNtwo$.  Contrast this situation with the result in Theorem \ref{the:Ma-Mi}(i) where the incentive strategy in $\PNtwo$ depends on the number of minor followers $N$ and employs deterministic strategies among all players with affine strategies in the action of the major follower. In addition, we note that if an approximate randomized mean-field incentive equilibrium exists, the preceding theorem does not require convexity and any differentiability condition on the cost function of the leader and the cost function of the major follower, compared to Theorem \ref{the:Ma-Mi}(i), where these conditions are required.

\section{Conclusions and Future Research Directions}
\label{sec:conc}
In this paper, we have studied stochastic incentive Stackelberg games with a dynamic information structure, where a leader seeks to elicit her desired response from a finite/infinite population of followers. We established the existence of an incentive strategy that attains the leader's desired performance and it sustains a symmetric  Stackelberg equilibrium with finitely many followers. Then, we proved that such strategies are not well-defined in a setting with infinitely many followers. In other words, the mean-field limit of such games is ill-defined. Then, we introduced a game variant, where the leader only incentivizes a major follower, who influences the other minor followers, who react to the major player's strategy through a Nash response. For this class of games, we established that, if incentive strategies converge in the infinite population limit, then they converge to an incentive strategy for the mean-field stochastic Stackelberg game. We further characterized the existence of an approximate mean-field incentive strategy for the class of stochastic Stackelberg mean-field games. Finally, we showed that a mean-field incentive strategy provides an approximation of an incentive strategy for the corresponding game with a finite but large number of minor followers.  

In this paper, we have only studied a single-stage interaction between the leader and the followers. A multi-stage dynamic game setting is a future direction of study that we aim to pursue. We also want to study randomized monitoring strategies in the multi-stage setting where the leader only monitors a subset of the times when interactions between the leader and the followers take place. Finally, we want to understand a possible way to merge the notion of information design with incentive design--a direction that merges Bayesian persuasion with incentive policy design for principal-agent settings.

\appendix
\section{Proof of Theorem \ref{the:symmetric}.}\label{sec:B}
 \begin{itemize}[wide]
\item [Part (i):] We use a technique similar to that used in \cite[Theorem 2.7]{sanjari2019optimal}. Let $\widehat{P}$ be the joint distribution of $y^{1:N}$ conditioned on $(y^{0}, \omega_{0})$, and $\widehat{P}^{0}$ be the joint distribution of $(y^{0}, \omega_{0})$. For every permutation $\sigma$ of $1, \dots, N$, we get
\begin{subequations}
\begin{flalign}
&J^{0}_{N}(\gamma^{0}_{\sigma},\gamma^{\sigma(1)}, \dots, \gamma^{\sigma(N)})\nonumber\\
&=\int c^{0}(\omega_{0}, u^{0}, \Lambda(u^{1}, \dots, u^{N})) \mathbbm{1}_{\{\gamma^{0}(y^{0}, y^{\sigma(1)}, \dots, y^{\sigma(N)}, u^{\sigma(1)}, \dots, u^{\sigma(N)})\in du^{0}\}}  
\nonumber
\\
& \quad \times \prod_{i=1}^{N}\mathbbm{1}_{\{\gamma^{\sigma(i)}(y^{i})\in du^{i}\}} \widehat{P}(dy^{1},\dots, dy^{N}\vert y^{0}, \omega_{0})\widehat{P}^{0}(dy^{0}, d\omega_{0})
\nonumber
\\
&=\int c^{0}(\omega_{0}, u^{0}, \Lambda(u^{\sigma(1)}, \dots, u^{\sigma(N)})) \mathbbm{1}_{\{\gamma^{0}(y^{0}, y^{1}, \dots, y^{N}, u^{1}, \dots, u^{N})\in du^{0}\}} 
\nonumber
\\
& \quad \times  \prod_{i=1}^{N}\mathbbm{1}_{\{\gamma^{\sigma(i)}(y^{\sigma(i)})\in du^{\sigma(i)}\}} \widehat{P}(dy^{\sigma(1)},\dots, dy^{\sigma(N)}\vert y^{0}, \omega_{0})\widehat{P}^{0}(dy^{0}, d\omega_{0})
\label{eq:p2}
\\
&=J^{0}_{N}(\gamma^{0}, \gamma^{1}, \dots, \gamma^{N}),\label{eq:p3}
\end{flalign}
\end{subequations}
where $\mathbbm{1}_{\{\cdot\}}$ denotes the indicator function, and $\gamma^{0}_{\sigma}(y^{0}, y^{1:N}, u^{1:N}):=\gamma^{0}(y^{0}, y^{\sigma(1):\sigma(N)}, u^{\sigma(1):\sigma(N)})$. Equality \eqref{eq:p2} follows from relabeling $y^{i}$ and $u^{i}$ with $y^{\sigma(i)}$ and $u^{\sigma(i)}$, respectively, and  \eqref{eq:p3} follows from the permutation invariant property of $\Lambda$ in \eqref{eq:lambda}, and exchangeablity of $y^{1:N}$ conditioned on $y^{0}, \omega_{0}$ in Assumption \ref{eq:syme}(iii).

Under the leader's information structure $I_{\CS}^{0}$ (or $I_{\PCS}^{0}$), and by convexity of $c^{0}$ in $u^{0:N}$ in Assumption \ref{eq:syme}(ii), \cite{SBSYateamsstaticreduction2021} yields that $J^{0}_{N}$ is convex in strategies\footnote{See \cite[Definition 3.1]{YukselSaldiSICON17} for the definition of convexity in strategies for teams.}. Denote the set of all permutations of $\{1,\dots, N\}$ by $S_{N}$ and its cardinality by $\vert S_{N}\vert$. Given any strategy profile $\gamma^{0:N}$ (possibly asymmetric), we have 
\begin{subequations}
\begin{flalign}
J^{0}_{N}(\gamma^{0:N})&=\frac{1}{\vert S_{N}\vert}\sum_{\sigma \in S_{N}} J^{0}_{N}(\gamma^{0}_{\sigma}, \gamma^{\sigma(1)}, \dots, \gamma^{\sigma(N)})\label{eq:p4}\\
&\geq J^{0}_{N}\left(\frac{1}{\vert S_{N}\vert}\sum_{\sigma \in S_{N}}\gamma^{0}_{\sigma}, \frac{1}{\vert S_{N}\vert}\sum_{\sigma \in S_{N}}\gamma^{\sigma(1)}, \dots, \frac{1}{\vert S_{N}\vert}\sum_{\sigma \in S_{N}}\gamma^{\sigma(N)}\right)\label{eq:p5}\\
&=J^{0}_{N}(\hat{\gamma}^{0}, \gamma, \dots, \gamma)\nonumber,
\end{flalign}
\end{subequations}
where \eqref{eq:p4} holds because the expected cost function of the leader is invariant to permutation of strategies, and \eqref{eq:p5} follows from Jensen's inequality. This completes the proof of Part (i) since the cost is bounded from below, an hence, an approximate optimal solution always exists. 

\item [Part (ii):] The proof proceeds in three steps. In Step 1, we show that there exists an ${\epsilon}^{0}$-leader-optimal strategy profile with the leader's strategy continuous in $u^{1:N}$. In Step 2, using Part (i) and Kakutani–Fan–Glicksberg fixed point theorem \cite[Corollary 17.55]{InfiniteDimensionalAnalysis}, we show that there exists a symmetric ${\epsilon}^{0}$-leader-optimal strategy profile with leader's strategy continuous in $u^{1:N}$ that admits a symmetric mixed Nash best response. By a mixed strategy, we mean that each follower independently randomizes their control actions via a stochastic kernel $\pi^{i}$ on $\mathbb{U}$, given $y^{i}$. Finally, in Step 3, we use a denseness argument similar to that used in \cite[Theorem 3]{milgrom1985distributional} to approximate said symmetric mixed Nash best response of the followers with a symmetric pure $\hat\epsilon$-Nash best response. 

We need additional notation for our proof. For any $\hat{\epsilon}\geq0$, denote the set of all $\gamma^{0}$'s of the leader with $R^{\hat{\epsilon}}(\gamma^{0})\cap \Gamma^{\SYM}\not=\emptyset$ by $\Gamma^{0, \SYMA}_{\hat{\epsilon}}$.  Define the space $\pi^{i}$ by $\Delta^{i}$ for each follower $i$.  Let $R^{\hat{\epsilon}}_{\pi}(\gamma^0),\Delta^{\SYM}$, and $\Gamma^{0, \SYMA}_{\hat{\epsilon}, \pi}$ define mixed strategy counterparts of various sets defined for pure strategy. 

\begin{itemize}[wide]
    \item [\textit{Step 1}.] Suppose that $(\gamma^{0\star}, \pi^{1\star:N\star})$ is an  $\bar{\epsilon}$-leader-optimal strategy profile. {An approximate leader-optimal strategy profile always exists since $c^{0}$ is bounded from below.} By Lusin's theorem, for every $\epsilon>0$, there exists a closed subset $C$ of $\prod_{i=1}^{N}\mathbb{U}$ such that $\mathbb{P}(\prod_{i=1}^{N}\mathbb{U}\backslash C)< \epsilon$, and the restriction of $\gamma^{0\star}_{y^{0:N}}$ to $C$ is continuous. By Tietze's extension theorem, there exists ${\gamma}^{0\star}_{c, y^{0:N}}$ continuous on $\prod_{i=1}^{N}\mathbb{U}$ such that ${\gamma}^{0\star}_{c, y^{0:N}}(u^{1:N})={\gamma}^{0\star}_{ y^{0:N}}(u^{1:N})$ on $C$, and hence, we have
    \begin{subequations}
    \begin{flalign}
    \bigg\vert\mathbb{E}^{\gamma^{0\star}, \pi^{1\star:N\star}}&\left[c^{0}(\omega_{0}, u^{0}, \Lambda(u^{1:N})\right]-\mathbb{E}^{\gamma^{0\star}_{c}, \pi^{1\star:N\star}}\left[c^{0}(\omega_{0}, u^{0}, \Lambda(u^{1:N})\right]\bigg\vert\nonumber
    \\
    &= \bigg\vert\int_{\prod_{i=1}^{N}\mathbb{U}\backslash C} g(\omega_{0}, \gamma^{0\star}_{y^{0:N}}(u^{1:N}), u^{1:N}, y^{0:N}) \prod_{i=1}^{N}\pi^{i\star}(du^{i}\vert y^{i})
    \nonumber
    \\
    &\qquad - \int_{\prod_{i=1}^{N}\mathbb{U}\backslash C} g(\omega_{0}, \gamma^{0\star}_{c, y^{0:N}}(u^{1:N}), u^{1:N}, y^{0:N}) \prod_{i=1}^{N}\pi^{i\star}(du^{i}\vert y^{i})\bigg\vert
    \label{ext}
    \\
    &\leq 2\|c^{0}\|_{\infty} \epsilon\label{eps},
    \end{flalign}
    \end{subequations}
    where 
    \begin{align}
        &g(\omega_{0}, \gamma^{0\star}_{y^{0:N}}(u^{1:N}), u^{1:N}, y^{0:N}):=\nonumber\\
        &\int_{\Omega_{0}\times \mathbb{Y}^{0} \times \prod_{i=1}^{N}\mathbb{Y}}c^{0}(\omega_{0}, \gamma^{0\star}_{y^{0:N}}(u^{1:N}), \Lambda(u^{1:N}))\lambda(d\omega_{0}, dy^{0:N}),
    \end{align}
    $\lambda$ is the joint distribution of $\omega_{0}, y^{0:N}$, and $\|c^{0}\|_{\infty}$ denotes a uniform bound for $c^{0}$. Therefore, $\gamma_{c}^{0\star}$ is $(2\|c^{0}\|_{\infty}+\bar{\epsilon})$-leader optimal strategy. 
    
    \item [\textit{Step 2}.] From part (i), there exists an approximate symmetric leader-optimal strategy profile, where followers' strategies are deterministic. Individual independent randomization by the followers cannot improve the leader's cost, according to  \cite[Theorem 2.1]{YukselSaldiSICON17}. Hence, that approximate symmetric leader-optimal strategy profile remains approximately optimal even if we allow for randomization, i.e., over $\Gamma^{0}\times \prod_{i=1}^{N} \Delta^{i}$. Further using Step 1, we can assume that the leader's strategy is continuous in the actions of followers. Call this leader's strategy $\gamma^{0}\in \Gamma^{0, \PI}$. 
Next, we show that $\gamma^{0}$ admits a symmetric mixed Nash best response strategy.

Under Assumption \ref{eq:symepart2}(ii), the expected cost of follower $i$ becomes
$J^i_N = \mathbb{E}\left[ \hat{c}(\tilde\omega_{0}, y^{i}, u^{i}, y^{-i}, u^{-i}) \right]$, where
 \begin{align}
     \hat{c}(\tilde\omega_{0}, y^{i}, u^{i}, y^{-i}, u^{-i}):={c}\left(\omega_{0},u^{i}, \gamma^{0}(y^{0:N}, u^{1:N}), \Lambda(u^{1:N})\right)\prod_{i=1}^{N}h(y^{i}, y^{0}, \omega_{0})
 \end{align} 
under a change of measure argument \cite[Definition 2.1]{SBSYateamsstaticreduction2021}. Here, we use the notation $\tilde\omega_{0}:=(\omega_{0},y^{0})$.
We now use Kakutani–Fan–Glicksberg fixed point theorem \cite[Corollary 17.55]{InfiniteDimensionalAnalysis} as follows. 
Define the reaction correspondence $\Phi:\prod_{i=1}^{N}\Delta^{i} \to 2^{\prod_{i=1}^{N}\Delta^{i}}$ as $\Phi(\hat\pi, \dots, \hat\pi):=\prod_{i=1}^{N} \text{BR}(\hat{\pi}^{-i})$, where
\begin{flalign}
&\text{BR}(\hat{\pi}^{-i}):=\bigg\{\pi^{i\star}\in \Delta^{i}\bigg\vert \nonumber\\
& \mathbb{E}^{\hat{\pi}^{-i}, \pi^{i}}\left[\hat{c}\left(\tilde\omega_{0}, y^{i}, u^{i}, y^{-i}, u^{-i}\right)\right]\geq \mathbb{E}^{\hat{\pi}^{-i}, \pi^{i\star}}\left[\hat{c}\left(\tilde\omega_{0}, y^{i}, u^{i}, y^{-i}, u^{-i}\right)\right]\:\:\forall \pi^{i}\in\Delta^{i}\bigg\}
\end{flalign}
for $i=1, \dots, N$.
Under the new measure on observation per Assumption \ref{eq:symepart2}(ii), $y^{1:N}$ are independent of each other and $\tilde\omega_{0}$. Hence, we can identify $\mathbb{P}$-almost surely, every mixed strategy of follower $i$ as a probability measure on $\mathbb{Y}\times \mathbb{U}$ with a fixed marginal $\tilde{\nu}$ on $\mathbb{Y}$. As a result, the set of all mixed strategies $\Delta^{i}$ of follower $i$ is convex and non-empty. Next, we show that $\text{BR}$ is upper hemicontinuous. Let a sequence of strategies of all followers  $\{\pi_{k}\}_{k}$  converge weakly to $\hat{\pi}$, and let $\pi_{k}\in \text{BR}(\pi_{k}^{-i})$.  Since ${\Lambda}$, $\gamma^{0}$, and $c$ are continuous in $u^{1:N}$, we get 
\begin{align}
\begin{aligned}
&\inf_{\pi^{i}\in \Delta^{i}}\mathbb{E}^{\hat{\pi}^{-i}, \pi^{i}}\left[\hat{c}\left(\tilde\omega_{0}, y^{i}, u^{i}, y^{-i}, u^{-i}\right)\right]\nonumber\\
&=\inf_{\pi^{i}\in \Delta^{i}}\lim_{k\to \infty}\mathbb{E}^{{\pi}^{-i}_{k}, \pi^{i}}\left[\hat{c}\left(\tilde\omega_{0}, y^{i}, u^{i}, y^{-i}, u^{-i}\right)\right]\\
&\geq \lim_{k\to \infty}\inf_{\pi^{i}\in \Delta^{i}}\mathbb{E}^{{\pi}^{-i}_{k}, \pi^{i}}\left[\hat{c}\left(\tilde\omega_{0}, y^{i}, u^{i}, y^{-i}, u^{-i}\right)\right]\\
&=\lim_{k\to \infty}\mathbb{E}^{{\pi}^{-i}_{k}, \pi^{i}_{k}}\left[\hat{c}\left(\tilde\omega_{0}, y^{i}, u^{i}, y^{-i}, u^{-i}\right)\right]\\
&=\mathbb{E}^{\hat{\pi}^{-i}, \hat{\pi}^{i}}\left[\hat{c}\left(\tilde\omega_{0}, y^{i}, u^{i}, y^{-i}, u^{-i}\right)\right].
\end{aligned}
\end{align}
In the above relation, the first and the last equalities follow from Assumptions \ref{eq:symepart2}(i) and \ref{eq:symepart2}(iii) using the generalized dominated convergence theorem \cite[Theorem 3.5]{serfozo1982convergence}. The second equality follows from $\pi_{k}\in \text{BR}(\pi_{k}^{-i})$.
Hence, $\hat{\pi}\in \text{BR}(\hat{\pi}^{-i})$, which implies that $\text{BR}$ is upper hemicontinuous. This implies that the graph 
\begin{align}
    G:=\left\{\left((\hat\pi, \dots, \hat\pi),  \Phi(\hat\pi, \dots, \hat\pi)\right)\bigg\vert (\hat\pi, \dots, \hat\pi) \in \prod_{i=1}^{N}\Delta^{i}\right\}
\end{align}
is closed. Also, since $\mathbb{U}$ is compact, $\Delta^{i}$ is compact under the $w$-$s$ topology, and by \cite[Proposition D.5(b)]{HernandezLermaMCP} the map 
\begin{align}
    F^{i}:\hat\pi^{-i} \mapsto \inf_{\pi^{i}\in \Delta^{i}}\mathbb{E}^{\hat{\pi}^{-i}, \pi^{i}}\left[\hat{c}\left(\tilde\omega_{0}, y^{i}, u^{i}, y^{-i}, u^{-i}\right)\right]
\end{align}
is
continuous in the $w$-$s$ topology. Since $\Delta^{i}$ is compact,  $\text{BR}(\hat\pi^{-i})$ is non-empty. Thus, by Kakutani–Fan–Glicksberg fixed point theorem \cite[Corollary 17.55]{InfiniteDimensionalAnalysis}, $\Phi$ admits a fixed point and this completes the proof of step 2.

\item [\textit{Step 3}.] In this step, we show that under Assumption \ref{eq:symepart2}(ii), for any leader strategy $\gamma^{0}\in \Gamma^{0, \PI}$  continuous in $u^{1:N}$, there exist symmetric pure $\hat{\epsilon}$-Nash best response strategies for the followers. By Part (ii), for any leader strategy $\gamma^{0}\in \Gamma^{0, \PI}$ continuous in $u^{1:N}$, there exists a symmetric mixed  Nash best response strategy for the followers. Since by Assumption \ref{eq:symepart2}(ii), the distribution of $y^{i}$, under the change of measure, is atomless and identical among followers, then appealing to \cite[Theorem 3]{milgrom1985distributional}, we get that $\Gamma^{i}$ is a dense subset of $\Delta^{i}$ in the $w$-$s$ topology. Since $\Gamma^{i}$ and $\Delta^{i}$ are identical among followers, following from the continuity of $F^i$, we can approximate the symmetric mixed Nash response arbitrarily closely using symmetric pure strategies. Owing to the continuity of costs, these deterministic strategies define approximate pure Nash responses from the followers.

\end{itemize}

\section{Proof of Theorem \ref{the:sym2}.}\label{sec:C}
Consider a symmetric $\epsilon_{0}$-leader-optimal strategy profile $(\tilde{\gamma}^{0}, \gamma^{\star}, \ldots, \gamma^{\star})$. By Assumption \ref{assump:the3.2}, the cost function $c$ of follower $i$ is strictly convex and continuously differentiable in $u^{i}$. Hence, the best response of follower $i$ is characterized by the following stationarity condition $\mathbb{P}$-a.s.,
\begin{flalign}
\begin{aligned}
\mathbb{E}\bigg[&\frac{\partial}{\partial u^{i}}c(\omega_{0}, \gamma^{0\star}(y^{0},y^{1:N}), u^{i}, \Lambda(\gamma^{-i\star}(y^{-i}), u^{i})) +Q(y^{i}, y^{0}, y^{-i})
\\
&\times\frac{\partial}{\partial u^{0}}c\left(\omega_{0}, u^{0}, \gamma^{\star}(y^{i}), \Lambda(\gamma^{\star}(y^{1}), \dots, \gamma^{\star}(y^{N}))\right)\bigg{\vert}y^{i}\bigg]_{\substack{{u^{i}=\gamma^{\star}(y^{i}),}\\{u^{0}=\gamma^{0\star}(y^{0},y^{1:N})}}}=0,
\end{aligned}
\label{eq:QQQNch3}
\end{flalign}
where $(\gamma^{-i\star}(y^{-i}), u^{i}):=(\gamma^{\star}(y^{1}), \dots,\gamma^{\star}(y^{i-1}) , u^{i}, \gamma^{\star}(y^{i+1}), \dots, \gamma^{\star}(y^{N}))$, as long as a Borel measurable $Q$ exists. Hahn-Banach Theorem gaurantees the existence of a bounded linear $Q$ that satisfies \eqref{eq:QQQNch3}. See \cite{bacsar1984affine} for details.

To show that $\left(\tilde{\gamma}^{0\star}, \gamma^{\star}, \ldots, \gamma^{\star}\right)$ constitutes a Stackelberg-incentive equilibrium, note that 
\begin{flalign}
\inf_{\gamma^{0}\in \Gamma^{0}_{\CS}} \inf _{\gamma^{1:N} \in R(\gamma^{0})} J^{0}_{N}(\gamma^{0:N})&\geq \inf_{\gamma^{0:N}\in \Gamma^{0}_{\CS}\times \prod_{i=1}^{N}\Gamma^{i}}  J^{0}_{N}( \gamma^{0:N})\\
&\geq J^{0}_{N}(\tilde{\gamma}^{0\star}, \gamma^{\star}, \dots, \gamma^{\star})-\epsilon_{0}.\label{eq:3leq3Nch3}
\end{flalign} 
The first inequality in \eqref{eq:3leq3Nch3} follows from 
the fact that  $R(\gamma^{0})\subseteq \prod_{i=1}^{N} \Gamma^{i}$. Any leader's strategy in $I_{\CS}^{0}$ that utilizes the controls of followers can be equivalently  represented as an explicit function of observations alone, i.e., within $I_{\NCS}^{0}$. Hence, infimizing $J^{0}_{N}( \gamma^{0:N})$ over $\gamma^{0:N}\in \Gamma^{0}_{\CS}\times \prod_{i=1}^{N}\Gamma^{i}$ is equivalent to infimizing the same over $\gamma^{0:N}\in \Gamma^{0}_{\NCS}\times \prod_{i=1}^{N}\Gamma^{i}$. The latter is, by definition, a leader-optimal strategy. Since $J^{0}_{N}(\tilde{\gamma}^{0\star}, \gamma^{\star}, \dots, \gamma^{\star}) = J^{0}_{N}({\gamma}^{0\star}, \gamma^{\star}, \dots, \gamma^{\star})
$, the second inequality in \eqref{eq:3leq3Nch3} follows. This completes the proof.

\section{Proof of Proposition \ref{the:neg1}.}\label{sec:D}

The leader-optimal solution can be written as a function of observations $y^{0:N}$ alone as ${\gamma}^{0\star}_N(y^{0:N})$. A differentiable incentive strategy therefore takes the form $g_{N}\left(y^{0:N},{\gamma}^{0\star}_{N}(y^{0:N}), \frac{1}{N}\sum_{p=1}^{N}u^{p}\right)$
with some measurable $g_N$ that is differentiable in its last argument.
It further satisfies
\begin{align}
    g_{N}\left(y^{0:N},{\gamma}^{0\star}_{N}(y^{0:N}), \frac{1}{N}\sum_{p=1}^{N}\gamma^{\star}_{N}(y^{p})\right)={\gamma}^{0\star}_{N}(y^{0:N}), \quad \mathbb{P}\text{-a.s.}
\end{align} 
since the leader-optimal solution for the QG game is unique, and is given by ${\gamma}^{0\star}_{N}(y^{0:N})=\alpha_{0}^{N} y^{0} + \frac{1}{N}\sum_{i=1}^{N} \alpha^{N}y^{i}$ and $\gamma^{\star}_{N}(y^{i})=\beta^{N}y^{i}$, where $\alpha^N_0, \alpha^N, \beta^N$ are given in \eqref{eq:QG1.ab}. In other words, since the response set of followers is unique, under the desired response from followers, the leader's optimal strategy will be unique up to a representation that coincides with ${\gamma}^{0\star}_{N}(y^{0:N})$. 
The stationarity criterion for incentive strategy dictates
\begin{align}
\begin{aligned}
&\mathbb{E}\bigg[r\beta^{N}y^{i}+ \left(1+\frac{1}{N}\right)q\left(\beta^{N}y^{i}+\alpha_{0}^{N} y^{0} +  \frac{1}{N}\sum_{p=1}^{N} \alpha^{N}y^{p}+ \omega_{0} +\frac{1}{N}\sum_{p=1}^{N} \beta^{N}y^{p}\right)
\\
&+\frac{q}{N}\frac{\partial g_N(y^{0:N},{\gamma}^{0\star}_{N}(y^{0:N}), z)}{\partial z} \nonumber\\
&\times\left(\beta^{N}y^{i}+ \alpha_{0}^{N} y^{0} + \frac{1}{N}\sum_{p=1}^{N} \alpha^{N}y^{p}+\omega_{0}+\frac{1}{N}\sum_{p=1}^{N} \beta^{N}y^{p}\right)\bigg\vert y^{i}\bigg]=0,
\end{aligned}
\end{align}
$\mathbb{P}$-a.s. for $i=1, \ldots, N$.
The above relation can be written as
\begin{flalign}
\mathbb{E}\left[\frac{\partial g_N}{\partial z}\kappa_{N}~\bigg\vert~y^{i}\right]=-\mathbb{E}\left[Nr\beta^{N}y^{i}+(N+1)\kappa_{N}~\bigg\vert~y^{i}\right], \quad \mathbb{P}\text{-a.s},\label{eq:Q77}
\end{flalign}
where $\kappa_{N}/q :=\beta^{N}y^{i}+\alpha_{0}^{N} y^{0} + \omega_{0} + \frac{1}{N}\sum_{p=1}^{N} \alpha^{N}y^{p}+\frac{1}{N}\sum_{p=1}^{N} \beta^{N}y^{p}$.
From \eqref{eq:QG1.ab}, $\alpha_{0}^{N}, \alpha^{N}, \beta^{N}$ converge as $N\to \infty$, and hence, $\lim_{N\to \infty}\mathbb{E}[\vert\kappa_{N}\vert^{2}\vert y^{i}]<\infty$ $\mathbb{P}$-a.s. for  $i=1, \dots, N$. Jensen's inequality yields
\begin{flalign}
\mathbb{E}\left[\frac{\partial g_N}{\partial z}\kappa_{N}\bigg\vert y^{i}\right]^{2}\leq \mathbb{E}\left[\left\vert\frac{\partial g_N}{\partial z}\kappa_{N}\right\vert^{2}\bigg\vert y^{i}\right] \quad \mathbb{P}\text{-a.s}
\label{eq:Q78}
\end{flalign}  
From \eqref{eq:Q77} and \eqref{eq:Q78}, we get $\mathbb{E}[\vert\frac{\partial g_N}{\partial z}\kappa_{N}\vert^{2}\vert y^{i}]$ grows unbounded as $N\to \infty$. From Cauchy-Schwarz inequality, we obtain
\begin{align}
&\mathbb{E}\left[\left\vert\frac{\partial g_N}{\partial z}\kappa_{N}\right\vert^{2}\bigg\vert y^{i}\right]\leq \mathbb{E}\left[\left\vert\frac{\partial g_N}{\partial z}\right\vert^{2}\bigg\vert y^{i}\right]\mathbb{E}\left[\left\vert\kappa_{N}\right\vert^{2}\vert y^{i}\right].
\end{align}
Boundedness of $\mathbb{E}[\vert\kappa_N\vert^2\vert y^{i}]$ implies that $\mathbb{E}[\vert\frac{\partial g_N}{\partial z}\vert^{2}\vert y^{i}]$ grows unbounded as $N\to \infty$. Then, Fatou's Lemma gives
\begin{align}
\liminf_{N\to \infty}\mathbb{E}\left[\mathbb{E}\left[\left\vert\frac{\partial g_N}{\partial z}\right\vert^{2}\bigg\vert y^{i}\right]\right]&\geq \mathbb{E}\left[\liminf_{N\to \infty}\mathbb{E}\left[\left\vert\frac{\partial g_N}{\partial z}\right\vert^{2}\bigg\vert y^{i}\right]\right],
\end{align}
that in turn, yields $\mathbb{E}[\vert\frac{\partial g_N}{\partial z}\vert^{2}]\to \infty$ as $N\to \infty$, completing the proof.
\end{itemize}

\section{Proof of Theorem \ref{the:Ma-Mi}.}\label{sec:E}
We prove the two parts separately. 

Part (i):
Given the strategies of the leader and the minor followers, $c^{M}$ remains strictly convex in  $u^{M}$ by Assumption \ref{assumption:m-m3l}(i).  Then,  Hahn-Banach Theorem and an argument similar to that used for the proof of Theorem \ref{the:sym2} yields the existence of $Q_{N}$ for which $(\tilde{\gamma
}^{0\star}_{N}, \gamma^{M\star}_{N}, \gamma^{1\star:N\star}_{N})$ satisfies
\begin{flalign}
\inf_{\gamma^{M}\in \Gamma^{M}}J^{M}_{N}(\tilde{\gamma
}^{0\star}_{N}, \gamma^{M}, \gamma^{1\star:N\star}_{N})=J^{M}_{N}(\tilde{\gamma
}^{0\star}_{N}, \gamma^{M\star}_{N}, \gamma^{1\star:N\star}_{N})\label{eq:step2}.
\end{flalign}
Since 
$\gamma^{1\star:N\star}_{N}\in R_{\ma}^{\hat{\epsilon}}(\gamma^{M\star}_{N})$ and $(\gamma^{M\star}_{N}, \gamma^{1\star:N\star}_{N})\in R^{\hat{\epsilon}}(\tilde{\gamma
}^{0\star}_{N})$, we infer 
\begin{subequations}
\begin{flalign}
&\inf_{\gamma^{0}\in \Gamma_{\MCS}^{0}} \inf_{(\gamma^{M}, \gamma^{1:N})\in R^{\hat{\epsilon}}(\gamma^{0})} J_{N}^{0}(\gamma^{0}, \gamma^{M}, \gamma^{1:N})\nonumber\\
&\geq \inf_{\gamma^{0}, \gamma^{M}\in \Gamma_{\MCS}^{0}\times \Gamma^{M}}  \inf_{\gamma^{1:N}\in R_{\ma}^{\hat{\epsilon}}(\gamma^{M})} J_{N}^{0}(\gamma^{0}, \gamma^{M}, \gamma^{1:N})
\label{eq:eq1}
\\
& = \inf_{\gamma^{0}, \gamma^{M}\in \Gamma_{\NCS}^{0}\times \Gamma^{M}}  \inf_{\gamma^{1:N}\in R_{\ma}^{\hat{\epsilon}}(\gamma^{M})} J_{N}^{0}(\gamma^{0}, \gamma^{M}, \gamma^{1:N})
\label{eq:eq2}
\\
& \geq  J_{N}^{0}(\tilde{\gamma}^{0\star}_{N}, \gamma^{M\star}_{N}, \gamma^{1\star:N\star}_{N}) - \epsilon_{0}.
\label{eq:eq3}
\end{flalign} 
\end{subequations}
Here, \eqref{eq:eq1} follows from the additional infimization over $\Gamma^{M}$ on the right-hand-side. The leader's strategy does not depend on the minor followers' actions; also, the minor followers' costs do not depend directly on the leader's actions. Therefore, a leader's strategy in $\Gamma^0_\MCS$ admits an equivalent representation in $\Gamma^0_\NCS$, for which the response of the minor followers remains unaltered. This observation leads to the conclusion in \eqref{eq:eq2}.
The relation in \eqref{eq:eq3} follows from the approximate leader-major optimality of $({\gamma}^{0\star}_{N}, \gamma^{M\star}_{N}, \gamma^{1\star:N\star}_{N})$ whose performance matches that of $(\tilde{\gamma}^{0\star}_{N}, \gamma^{M\star}_{N}, \gamma^{1\star:N\star}_{N})$. 

Part (ii): This proof proceeds in three steps. In step 1, we show that the minor followers' strategies $(\gamma^{\star}_{\infty}, \mu^{\star})$ is a Nash best response to $\gamma^{M\star}_{\infty}$. In step 2, we show that the strategy profile $(\tilde{\gamma}^{0\star}_{\infty}, \gamma^{M\star}_{\infty}, \gamma^{\star}_{\infty})$ constitutes an ${\epsilon}$-leader-major mean-field optimal  for $\Pitwo$. Finally, in step 3, we show that the strategy profile $(\tilde{\gamma}^{0\star}_{\infty}, \gamma^{M\star}_{\infty}, \gamma^{\star}_{\infty})$ defines an approximate Stackelberg mean-field equilibrium. 
\begin{itemize}[wide]
\item [{\it Step 1.}] We prove that $(\gamma^{\star}_{\infty}, \mu^{\star})\in R^{\infty, \hat{\epsilon}}_{\ma}(\gamma^{M\star}_{\infty})$. To that end, consider a Borel set $\mathbb{A} \subseteq \mathbb{U}\times \mathbb{Y}$, over which the empirical measure $e_{N}\in \mathcal{P}(\mathbb{U}\times \mathbb{Y})$ is given by
\begin{flalign}
&e_{N}(\mathbb{A}):=\frac{1}{N}\sum_{i=1}^{N}\delta_{\{(\gamma^{\star}_{N}(y^{i}, y^{M}), y^{i})\}}(\mathbb{A})\label{emp},
\end{flalign}
where $\delta$ defines the Dirac-Delta measure. Assumptions \ref{assumption:m-m3l}(iii) and \ref{assumption:m-m3l}(v) allow a change of measure, under which $y^{1:N}$ are i.i.d., given $y^M$. Also, the sequence of strategies $\{\gamma^{\star}_{N}\}_{N}$ converges point-wise to $\gamma^{\star}_{\infty}$. 
Then, using the strong law of large numbers, we conclude that $\{e_{N}\}_{N}$ converges to $e_{\infty}:=\mathcal{L}(\gamma^{\star}_{\infty}(y^{R}, y^{M})\vert y^{M})\tilde\nu(dy^{R})$ $\mathbb{Q}$-a.s. in the $w$-$s$ topology due to Glivencko-Cantelli Theorem, where $\mathbb{Q}:=\mathbb{P}^{0}\times \tilde{\nu}^{M}\times \tilde\nu^{0}$ and $\mathbb{P}^{0}$ is the distribution of $\omega
_{0}$. Thanks to Assumption \ref{assumption:m-m3l}(v), $e_{\infty}$ is independent of $\omega_{0}$. Given that $\mathbb{U}$ is compact, we have 
\begin{flalign}
\lim\limits_{N\to \infty}\int_{\mathbb{U}} u e_{N}(du\times \mathbb{Y}) = \int_{\mathbb{U}} u e_{\infty}(du\times \mathbb{Y}).
\quad \mathbb{Q}\text{-a.s.},
\label{eq:empconverg}
\end{flalign}
conditioned on $y^{M}$.
Next, we have
\begin{align}
\begin{aligned}
&\limsup_{N\to \infty}\ \mathbb{E} \left[\mathbb{E}\left[c\left(\omega_{0}, \gamma^{\star}_{N}(y^{i}, y^{M}), \gamma^{M\star}_{N}(y^{M}), \frac{1}{N}\sum_{p=1}^{N}\gamma^{\star}_{N}(y^{p}, y^{M})\right)\bigg\vert y^{M}\right]\right]
\\
&\quad =\mathbb{E}\left[c\left(\omega_{0}, \gamma^{\star}_{\infty}(y^{i}, y^{M}), \gamma^{M\star}_{\infty}(y^{M}), \int_{\mathbb{U}} u e_{\infty}(du\times \mathbb{Y})\right)\right],
\end{aligned}
\label{eq:converg}
\end{align}
using \eqref{eq:empconverg}, the law of total expectation, the generalized dominated convergence theorem \cite[Theorem 3.5]{serfozo1982convergence}, point-wise convergence of $\{\gamma^{M\star}_{N}\}_{N}$ and $\{\gamma^{\star}_{N}\}_{N}$ to $\gamma^{M\star}_{\infty}$ and $\gamma^{\star}_{\infty}$, respectively, and Assumptions \ref{assumption:m-m3l}(ii) and \ref{assumption:m-m3l}(vi). Then, we infer
\begin{subequations}
\begin{flalign}
J^{R}_{\infty}(\gamma^{M\star}_{\infty}, \gamma^{\star}_{\infty}, \mu^{\star}) &= \limsup_{N\to \infty} \ J^{R}_{N}(\gamma^{M\star}_{N}, \gamma^{\star}_{N}, \dots, \gamma^{\star}_{N})\label{eq:L1}\\
& \leq \limsup_{N\to \infty} \inf_{\gamma^{R}\in \Gamma^{R}}J^{R}_{N}(\gamma^{M\star}_{N}, \gamma^{-R\star}_{N}, \gamma^{R})+\hat{\epsilon}\label{eq:L2}\\
& \leq \inf_{\gamma^{R}\in \Gamma^{R}} \limsup_{N\to \infty} \ J^{R}_{N}(\gamma^{M\star}_{N}, \gamma^{-R\star}_{N}, \gamma^{R})+\hat{\epsilon}\label{eq:xL2}\\
& = \inf_{\gamma^{R}\in \Gamma^{R}} J^{R}_{\infty}(\gamma^{M\star}_{\infty}, \gamma^{R}, \mu^{\star})+\hat{\epsilon}\label{eq:L3}.
\end{flalign}
\end{subequations}
Here, \eqref{eq:L1} follows from \eqref{eq:converg}, and \eqref{eq:L2} from  $(\gamma^{\star}_{N}, \dots, \gamma^{\star}_{N})\in R_{\ma}^{\hat{\epsilon}}(\gamma^{M\star}_{N})$ for every $N\in \mathbb{N}$. 
We obtain \eqref{eq:L3} from an argument similar to the derivation of \eqref{eq:converg} to get
\begin{align}
\begin{aligned}
&\limsup_{N\to \infty}\  J^{R}_{N}(\gamma^{M\star}_{N}, \gamma^{-R\star}_{N}, \gamma^{R})
\\
&=\limsup_{N\to \infty}\nonumber\\
\  &\mathbb{E}\left[c\left(\omega_{0}, \gamma^{R}(y^{R}, y^{M}),\gamma^{M\star}_{N}(y^{M}),\frac{1}{N}\left[\sum_{i=1, i\not= R}^{N}\gamma^{\star}_{N}(y^{i}, y^{M})+\gamma^{R}(y^{R}, y^{M})\right]\right)\right]\\
&=\mathbb{E}\left[c\left(\omega_{0}, \gamma^{R}(y^{R}, y^{M}),\gamma^{M\star}_{\infty}(y^{M}),\int u e_{\infty}(du\times \mathbb{Y})\right)\right],
\end{aligned}
\end{align}
since $\mathbb{U}$ is compact, and that actions taken via $\gamma^{\star}_{\infty}$ generate $\mu^{\star}=e_{\infty}(\cdot \times \mathbb{Y})$. This completes the proof of step 1. 

\item [{\it Step 2.}] In this step, we show that $(\tilde{\gamma}^{0\star}_{\infty}, \gamma^{M\star}_{\infty}, \gamma^{\star}_{\infty})$ constitutes an ${\epsilon}$-leader-major mean-field optimal  for $\Pitwo$.
We have
\begin{subequations}
\begin{flalign}
&J_{\infty}^{0}(\tilde{\gamma}^{0\star}_{\infty}, \gamma^{M\star}_{\infty}, \gamma^{\star}_{\infty})\nonumber\\
&= \limsup_{N\to \infty}\  J_{N}^{0}(\tilde{\gamma}^{0\star}_{N}, \gamma^{M\star}_{N}, \gamma^{\star}_{N}, \dots, \gamma^{\star}_{N})\label{eq:L5}\\
& \leq \limsup_{N\to \infty}\  \inf_{\gamma^{0}, \gamma^{M}\in \Gamma_{\MCS}^{0}\times \Gamma^{M}}  \inf_{\gamma^{1:N}\in R_{\ma}^{\hat{\epsilon}}(\gamma^{M}) \cap \Gamma^{\SYM}} J_{N}^{0}(\gamma^{0}, \gamma^{M}, \gamma^{1:N})+\epsilon_{0}\label{eq:L7}\\
& \leq  \inf_{\gamma^{0}, \gamma^{M}\in \Gamma_{\MCS}^{0}\times \Gamma^{M}}  \limsup_{N\to \infty}\  \inf_{\gamma^{1:N}\in R_{\ma}^{\hat{\epsilon}}(\gamma^{M}) \cap \Gamma^{\SYM}} J_{N}^{0}(\gamma^{0}, \gamma^{M}, \gamma^{1:N})+\epsilon_{0}\label{eq:ccL7}\\
&= \inf_{\gamma^{0}, \gamma^{M}\in \Gamma_{\MCS}^{0}\times \Gamma^{M}}  \inf_{(\gamma^{R}, \mu)\in R^{\infty, \hat{\epsilon}}_{\ma}(\gamma^{M})} J_{\infty}^{0}(\gamma^{0}, \gamma^{M}, \gamma^{R})+\epsilon_{0}.
\label{eq:L8}
\end{flalign}
\end{subequations}
The derivation of \eqref{eq:L5} follows an argument similar to that used to establish \eqref{eq:converg}. The inequality in  \eqref{eq:L7} follows from part (i) and that $(\tilde{\gamma}^{0\star}_{N}, \gamma^{M\star}_{N}, \gamma^{\star}_{N}, \dots, \gamma^{\star}_{N})$ defines an ${\epsilon}$-leader-major optimal solution for $\PNtwo$ with symmetric response from the minor followers for all $N\in \mathbb{N}$. 
Derivation of \eqref{eq:L8} from \eqref{eq:ccL7} follows from Assumptions \ref{assumption:m-m3l}(ii) and \ref{assumption:m-m3l}(vi), along the lines of that used to establish \eqref{eq:converg}; details are omitted for brevity. Step 2 then follows from step 1, given  $(\gamma^{\star}_{\infty}, \mu^{\star})\in R^{\infty, \hat{\epsilon}}_{\ma}(\gamma^{M\star}_{\infty})$.

\item [{\it Step 3.}] In this step, we show that $(\tilde{\gamma}^{0\star}_{\infty}, \gamma^{M\star}_{\infty}, \gamma^{\star}_{\infty})$ constitutes an approximate Stackelberg mean-field equilibrium. Along the lines of step 2, we infer
\begin{subequations}
\begin{flalign}
J^{M}_{\infty}(\tilde{\gamma}^{0\star}_{\infty}, \gamma^{M\star}_{\infty}, \gamma^{\star}_{\infty}) 
&= \limsup_{N\to \infty}\  \inf_{\gamma^{M}\in \Gamma^{M}} J^{M}_{N}(\tilde{\gamma}^{0\star}_{N}, \gamma^{M}, \gamma^{\star}_{N}, \dots,  \gamma^{\star}_{N})\label{eq:L11}\\
&\leq  \inf_{\gamma^{M}\in \Gamma^{M}} \limsup_{N\to \infty}\  J^{M}_{N}(\tilde{\gamma}^{0\star}_{N}, \gamma^{M}, \gamma^{\star}_{N}, \dots,  \gamma^{\star}_{N})\label{eq:oL12}\\
&= \inf_{\gamma^{M}\in \Gamma^{M}}J^{M}_{\infty}(\tilde{\gamma}^{0\star}_{\infty}, \gamma^{M}, \gamma^{\star}_{\infty})\label{eq:L12},
\end{flalign}
\end{subequations}
where \eqref{eq:L11} and  \eqref{eq:L12} follow from \eqref{eq:step2} and an argument similar to that used to establish \eqref{eq:converg}; the details are omitted. From the above relation, we obtain
\begin{align}
\inf_{\gamma^{M}\in \Gamma^{M}}J^{M}_{\infty}(\tilde{\gamma}^{0\star}_{\infty}, \gamma^{M}, \gamma^{\star}_{\infty})=J^{M}_{\infty}(\tilde{\gamma}^{0\star}_{\infty}, \gamma^{M\star}_{\infty}, \gamma^{\star}_{\infty}),
\label{eq:L9}
\end{align}
and hence, by step 1, we have $(\gamma^{M\star}_{\infty}, \gamma^{\star}_{\infty}, \mu^{\star})\in R^{\infty, \hat{\epsilon}}(\tilde{\gamma}^{0\star}_{\infty})$. Using the approximate leader-major mean-field optimality of $(\tilde{\gamma}^{0\star}_{\infty}, \gamma^{M\star}_{\infty}, \gamma^{\star}_{\infty})$ in
\begin{align}
\begin{aligned}
&\inf_{\gamma^{0}\in \Gamma_{\MCS}^{0}} \inf_{(\gamma^{M}, \gamma^{R}, \mu) \in R^{\infty, \hat{\epsilon}}(\gamma^{0})} J^{0}_{\infty}(\gamma^{0}, \gamma^{M}, \gamma^{R})\nonumber\\
&\geq \inf_{(\gamma^{0}, \gamma^{M}) \in \Gamma_{\MCS}^{0}\times \Gamma^{M}} \inf_{(\gamma^{R}, \mu) \in R^{\infty, \hat{\epsilon}}_{\ma}(\gamma^{M})} J^{0}_{\infty}(\gamma^{0}, \gamma^{M}, \gamma^{R})
\\
&\geq  J_{\infty}^{0}(\tilde{\gamma}^{0\star}_{\infty}, \gamma^{M\star}_{\infty}, \gamma^{\star}_{\infty})-\epsilon_{0},
\end{aligned}
\label{eq:L13}
\end{align}
we conclude the proof of part (ii). The first inequality above follows from a similar argument as that used in \eqref{eq:eq1}.
\end{itemize}


\section{Proof of Proposition \ref{lem:exi-1}.}
\label{A:F}


Part (i): Using an argument similar to that used in step 2 of the proof of Theorem \ref{the:symmetric}(ii), $R_{\ma, \pi}(\pi^{M})$ is nonempty for each $\pi^{M}\in \Delta^{M}$. Given that $J^0_N$ is bounded below, this guarantees the existence of an $(\bar\epsilon^{0}, 0)$-leader-major optimal solution $(\pi^{0\star}, \pi^{M\star}, \pi^{1\star:N\star})\in \Delta^{0}_{\NCS} \times \Delta^{M}\times \prod_{i=1}^{N}\Delta^{i}$ for $\PNtwo$ for every $\bar\epsilon^{0}>0$. 

Assumption \ref{assum:5}(iv)-\ref{assum:5}(v) allows a change of measure under which the distribution of $y^{M}$ is atomless. Then, using \cite[Theorem 3]{milgrom1985distributional}, $\Gamma^{M}$ is dense in $\Delta^{M}$ in the $w$-$s$ topology. Hence, there exists a sequence of pure strategies $\gamma^{M\star}_{k}$ in $\Gamma^{M}$ that converges weakly to $\pi^{M\star}$. Next, we show that $R_{\ma, \pi}: \Delta^{M} \to \prod_{p=1}^{N}\Delta^{p}$ is an upper hemicontinuous correspondence. To that end, notice that if $\pi^{1:N}_{k}\in R_{\ma, \pi}(\pi^{M}_{k})$, and $\pi^{M}_{k}$ and $\pi^{p}_{k}$ converge weakly to $\pi^{M}$ and $\pi^{p}$, respectively, as $k\to \infty$ for $p=1, \ldots, N$, then
\begin{align}
\begin{aligned}
\inf_{\tilde\pi^{i}\in \Delta^{i}} J^{i}(\pi^{M}, \tilde\pi^{i}, \pi^{-i})&=\inf_{\tilde\pi^{i}\in \Delta^{i}} \lim_{k\to \infty}J^{i}(\pi^{M}_{k}, \tilde\pi^{i}, \pi^{-i}_{k})\\
& \geq \lim_{k\to \infty}\inf_{\tilde\pi^{i}\in \Delta^{i}} J^{i}(\pi^{M}_{k}, \tilde\pi^{i}, \pi^{-i}_{k})\\
&=\lim_{k\to \infty}J^{i}(\pi^{M}_{k}, \pi^{i}_{k}, \pi^{-i}_{k})\\
&= J^{i}(\pi^{M}, \pi^{i}, \pi^{-i}).
\end{aligned}
\end{align}
In the above, the first and the last equalities follow from Assumption \ref{assum:5}(ii) and the generalized dominated convergence theorem. The second equality follows from  $\pi^{1:N}_{k}\in R_{\ma, \pi}(\pi^{M}_{k})$. This shows that $R_{\ma, \pi}$ is upper hemicontinuous. Since $\mathbb{U}$ is compact, and the marginal on $\mathbb{Y}$ is fixed, $\Delta^{i}$ is tight. Also, $\Delta^{i}$ is  closed, which implies that $\Delta^{i}$ is compact. Since $J^{i}(\pi^{i}, \pi^{-i}, \pi^{M})$ is weakly continuous in $\pi^{i}$ for any $\pi^{-i}, \pi^{M}$, by \cite[Proposition D.5(b)]{HernandezLermaMCP}, the map $F^{i}:\pi^{M}\mapsto \inf_{\tilde\pi^{i}\in \Delta^{i}}J^{i}(\pi^{M},\tilde\pi^{i}, \pi^{-i})$ is continuous. Hence, there exists a pure strategy $\gamma^{M}_{\hat{\epsilon}}$ in $\Gamma^{M}$ such that
\begin{flalign}
J^{i}(\gamma^{M\star}_{\hat{\epsilon}},\pi^{i\star}, \pi^{-i\star})&\leq J^{i}(\pi^{M\star},\pi^{i\star}, \pi^{-i\star})+\bar\epsilon
 \nonumber\\
 &
=\inf_{\pi^{i}\in \Delta^{i}}J^{i}(\pi^{M\star},\pi^{i}, \pi^{-i\star})+\bar{\epsilon}
 \nonumber\\
 &
\leq 
\inf_{\pi^{i}\in \Delta^{i}}J^{i}(\gamma^{M}_{\hat{\epsilon}},\pi^{i}, \pi^{-i\star})+\hat{\epsilon}.
\end{flalign}
This implies that $\pi^{1\star:N\star}\in R_{\ma, \pi}^{\hat\epsilon}(\gamma^{M\star}_{\hat{\epsilon}})$.
By continuity of $J^0_N$ in $\pi^M$, we obtain that 
\begin{flalign}
&\left\vert J_{N}^{0}(\pi^{0\star}, \pi^{M\star}, \pi^{1\star:N\star})-J_{N}^{0}(\pi^{0\star}, \gamma^{M\star}_{\hat\epsilon}, \pi^{1\star:N\star})\right\vert\leq \tilde\epsilon^{0}\label{eq:lem105}.
\end{flalign}


Since $(\pi^{0\star}, \pi^{M\star}, \pi^{1\star:N\star})$ is $(\bar{\epsilon}^{0},0)$-leader-major optimal, \eqref{eq:lem105} implies that
\begin{flalign}
&J_{N}^{0}(\pi^{0\star}, \gamma^{M\star}_{\hat\epsilon}, \pi^{1\star:N\star})\nonumber\\
&\qquad -\inf_{\pi^{0}, \pi^{M}\in \Delta^{0}_{\NCS} \times \Delta^{M}} \inf_{\pi^{1:N}\in R_{\ma, \pi}(\pi^{M})} J^{0}_{N}(\pi^{0}, \pi^{M}, \pi^{1:N})\leq \tilde\epsilon^{0}+\bar{\epsilon}^{0}\label{eq:lem106}.
\end{flalign}



Since $\Delta^{0}_{\NCS}$ is convex and $J^0_N$ is linear in randomized policies, and the selection of $\pi^{0\star}$ does not directly affect $R_{\ma, \pi}^{\hat\epsilon}(\gamma^{M\star}_{\epsilon})$, there exists a pure leader's strategy $\gamma^{0\star}\in \Gamma^{0}_{\NCS}$  such that
\begin{flalign}
&J_{N}^{0}(\pi^{0\star}, \gamma^{M\star}_{\hat\epsilon}, \pi^{1\star:N\star})=J_{N}^{0}(\gamma^{0\star}, \gamma^{M\star}_{\hat\epsilon}, \pi^{1\star:N\star})\label{eq:lem108}.
\end{flalign}


 Adding and subtracting $\inf\limits_{\gamma^{0}, \gamma^{M}\in \Gamma^{0}_{\NCS} \times \Gamma^{M}} \inf\limits_{\pi^{1:N}\in R_{\ma, \pi}(\pi^{M})} J^{0}_{N}(\gamma^{0}, \gamma^{M}, \pi^{1:N})$ in \eqref{eq:lem106}, and using \eqref{eq:lem108}, we obtain
\begin{flalign*}
&\bigg(\inf_{\gamma^{0}, \gamma^{M}\in \Gamma^{0}_{\NCS} \times \Gamma^{M}} \inf_{\pi^{1:N}\in R_{\ma, \pi}^{\hat\epsilon}(\pi^{M})} J^{0}_{N}(\gamma^{0}, \gamma^{M}, \pi^{1:N})\\
&-\inf_{\pi^{0}, \pi^{M}\in \Delta^{0}_{\NCS} \times \Delta^{M}} \inf_{\pi^{1:N}\in R_{\ma, \pi}(\pi^{M})} J^{0}_{N}(\pi^{0}, \pi^{M}, \pi^{1:N})\bigg)\nonumber\\
&+\bigg(J_{N}^{0}(\gamma^{0\star}, \gamma^{M\star}_{\hat\epsilon}, \pi^{0\star:N\star}) \\
&-\inf_{\gamma^{0}, \gamma^{M}\in \Gamma^{0}_{\NCS} \times \Gamma^{M}} \inf_{\pi^{1:N}\in R_{\ma, \pi}^{\hat\epsilon}(\pi^{M})} J^{0}_{N}(\gamma^{0}, \gamma^{M}, \pi^{1:N})\bigg)\leq \tilde{\epsilon}^{0}+\bar{\epsilon}^{0}.
\end{flalign*}
Since $\pi^{1\star:N\star}\in R_{\ma, \pi}^{\hat\epsilon}(\gamma^{M\star}_{\hat{\epsilon}})$ and $c^{0}$ is bounded, the second expression in the above is non-negative and bounded, and this implies \eqref{eq:mainlem1} and completes the proof.

Part (ii): We have that $(\gamma^{0\star}, \gamma^{M\star}, \pi^{1\star:N\star})\in \Gamma^{0}_{\NCS}\times \Gamma^{M}\times \prod_{i=1}^{N}\Delta^{i}$  constitutes $\epsilon$-leader-major optimal solution for $\PNtwo$ among all strategies, belonging to $\Delta^{0}_{\NCS}\times \Delta^{M}\times \prod_{i=1}^{N}\Delta^{i}$. By Hahn Banach theorem and an argument used in the proof of Theorem \ref{the:Ma-Mi}(i) (see step 1), there exists  $\tilde{\gamma}^{0\star}\in \Gamma^{0}_{\MCS}$ of the form \eqref{eq:incentive3l} (based on strategies $\gamma^{0\star}$ and $\gamma^{M\star}$) such that $\inf_{\gamma^{M}\in \Gamma^{M}} J^{M}_{N}(\tilde{\gamma}^{0\star}, \gamma^{M}, \pi^{1\star:N\star})=J^{M}_{N}(\tilde{\gamma}^{0\star}, \gamma^{M\star}, \pi^{1\star:N\star})$.
Setting strategies of the leader and minor followers to $\tilde{\gamma}^{0}$ and $\pi^{1\star:N\star}$, respectively, in the cost function $c^{M}$ of the major follower, since $\Delta^{M}$ is convex and $J^M_N$ is linear in randomized policies, we have  
\begin{flalign}
\inf_{\gamma^{M}\in \Gamma^{M}} J^{M}_{N}(\tilde{\gamma}^{0\star}, \gamma^{M}, \pi^{1\star:N\star})=
\inf_{\pi^{M}\in \Delta^{M}} J^{M}_{N}(\tilde{\gamma}^{0\star}, \pi^{M}, \pi^{1\star:N\star})\label{eq:t112}.
\end{flalign}
Since $(\gamma^{0\star}, \gamma^{M\star}, \pi^{1\star:N\star})$ is an $\epsilon$-leader-major optimal solution, we have that $\pi^{1\star:N\star}\in R_{\ma, \pi}^{\hat{\epsilon}}(\gamma^{M\star})$, and hence \eqref{eq:t112} implies that $(\gamma^{M\star}, \pi^{1\star:N\star})\in R_{\pi}^{\hat\epsilon}(\tilde{\gamma}^{0\star})$. Hence, we have
\begin{subequations}
\begin{flalign}
&\inf_{\pi^{0}\in \Delta^{0}_{\MCS}} \inf_{(\pi^{M}, \pi^{1:N})\in R_{\pi}^{\hat\epsilon}(\pi^{0})}J^{0}_{N}(\pi^{0}, \pi^{M}, \pi^{1:N})\nonumber\\
&\geq \inf_{(\pi^{0}, \pi^{M})\in \Delta^{0}_{\MCS}\times \Delta^{M}} \inf_{\pi^{1:N}\in R_{\ma, \pi}^{\hat\epsilon}(\pi^{M})}J^{0}_{N}(\pi^{0}, \pi^{M}, \pi^{1:N})\label{eq:ex1}\\
& = \inf_{(\gamma^{0}, \pi^{M})\in \Gamma^{0}_{\MCS}\times \Delta^{M}} \inf_{\pi^{1:N}\in R_{\ma, \pi}^{\hat\epsilon}(\pi^{M})}J^{0}_{N}(\gamma^{0}, \pi^{M},\pi^{1:N})\label{eq:ex2}\\
&= \inf_{(\gamma^{0}, \pi^{M})\in \Gamma^{0}_{\NCS}\times \Delta^{M}} \inf_{\pi^{1:N}\in R_{\ma, \pi}^{\hat\epsilon}(\pi^{M})}J^{0}_{N}(\gamma^{0}, \pi^{M},\pi^{1:N})\label{eq:ex3}\\
& \geq J^{0}_{N}(\tilde{\gamma}^{0\star}, \gamma^{M\star}, \pi^{1\star:N\star})-\epsilon^{0}\label{eq:ex4},
\end{flalign}
\end{subequations}
where \eqref{eq:ex1} follows from the fact that in the right-hand side the infimum is over all randomized strategies $\Delta^{M}$, and \eqref{eq:ex2} follows from convexity of $\Delta^{0}_{\MCS}$ since the strategy of the leader does not influence $R_{\ma, \pi}^{\hat\epsilon}(\pi^{M})$. Equality \eqref{eq:ex3} follows from nestedness of the information structure of the leader, following from an argument similar to that used in the proof of Theorem \ref{the:Ma-Mi}, and \eqref{eq:ex4} follows from the fact that $(\gamma^{0\star}, \gamma^{M\star}, \pi^{1\star:N\star})$ is an $\epsilon$-leader-major optimal solution and $(\tilde\gamma^{0\star}, \gamma^{M\star}, \pi^{1\star:N\star})$ attains the identical expected cost for the leader. Since $(\gamma^{M\star}, \pi^{1\star:N\star})\in R_{\pi}^{\hat\epsilon}(\tilde{\gamma}^{0\star})$, then $(\tilde\gamma^{0\star}, \gamma^{M\star}, \pi^{1\star:N\star})$ constitutes an $\epsilon$-Stackelberg equilibrium for $\PNtwo$ among all strategies, belonging to $\Delta^{0}_{\MCS}\times \Delta^{M}\times \prod_{i=1}^{N}\Delta^{i}$. The proof is completed since $(\tilde\gamma^{0\star}, \gamma^{M\star}, \pi^{1\star:N\star})\in \Gamma^{0}_{\MCS}\times \Gamma^{M}\times \prod_{i=1}^{N}\Delta^{i}$.

\section{Proof of Theorem \ref{the:Existence}.}\label{A:H}

 Part (i): The proof of this part proceeds in three steps. In step 1, we show that the space of randomized strategies for players are compact in the $w$-$s$ topology. This allows us to claim existence of a subsequence of an approximate symmetric Stackelberg-incentive equilibrium $(\pi^{0\star}_{n}, \pi^{M\star}_{n}. \pi^{\star}_{n},\ldots, \pi^{\star}_{n})$ for $\PNtwo$ as $N \to \infty$. Such an equilibrium exists for each $N$, thanks to Proposition \ref{lem:exi-1}. The above subsequence converges weakly to a limit strategy profile $(\pi^{0\star}_{\infty}, \pi^{M\star}_{\infty}, \pi^{\star}_{\infty},\ldots)$ as $N\to \infty$. In step 2, we show that $\pi^{\star}_{\infty}$ is an approximate mean-field Nash best response to $\pi^{M\star}_{\infty}$. Similarly, in step 3, we show that $(\pi^{0\star}_{\infty}, \pi^{M\star}_{\infty}. \pi^{\star}_{\infty},\ldots)$ constitutes an approximate leader-major optimal solution for $\Pitwo$, and  $(\pi^{0\star}_{\infty}, \pi^{M\star}_{\infty}. \pi^{\star}_{\infty},\ldots)$ is an approximate mean-field incentive equilibrium for $\Pitwo$.

\begin{itemize}[wide]
    \item [{\it Step 1.}] Under Assumption \ref{assum:5}(v), a change of measure argument allows us to define the space of randomized strategies for the leader and the major follower with independent randomization as
    $\hat{\nu}:= \tilde\nu^{0} \times \tilde\nu^{M}$ on $\mathbb{Y}^{0}\times \mathbb{Y}^{M}$ as 
    \begin{flalign}
    \begin{aligned}
    \Delta^{{0, M}}(\hat{\nu}):=\bigg\{&P \in \mathcal{P}(\mathbb{U}^{0}\times \mathbb{U}^{M}\times \mathbb{Y}^{0} \times \mathbb{U}^{M}) \bigg\vert~ \forall \mathbb{A} \in \mathcal{B}(\mathbb{U}^{0}\times \mathbb{U}^{M}\times \mathbb{Y}^{0} \times \mathbb{U}^{M})
    \\
    & P(\mathbb{A}) = \int_{\mathbb{A}} \pi^{0}(du^{0}\vert y^{0}, y^{M}, u^{M})\tilde{\nu}^{0}(dy^{0})\pi^{M}(du^{M}\vert y^{M})\tilde{\nu}^{M}( dy^{M})\bigg\},
    \end{aligned}
    \end{flalign}
    equipped with the $w$-$s$ topology. 
    Under Assumption \ref{assum:5}(v), we  also identify the set of randomized strategies of the major and minor followers, respectively, as 
    \begin{align}
    \begin{aligned}
    \Delta^{M}(\tilde\nu^{M})&:=\bigg\{P^{M} \in \mathcal{P}(\mathbb{U}^{M}\times \mathbb{Y}^{M})\bigg\vert~ \forall~\mathbb{A}\in \mathcal{B}(\mathbb{U}^{M}\times \mathbb{Y}^{M}) \\
    &\qquad \quad P^{M}(\mathbb{A})=\int_{\mathbb{A}} \pi^{M}(du^{M}\vert y^{M}) \tilde\nu^{M}(dy^{M})\bigg\},
    \\
    \Delta^{i}(\tilde\nu)&:=\left\{P^{i} \in \mathcal{P}(\mathbb{U}\times \mathbb{Y})\bigg\vert~ P^{i}(\mathbb{A})=\int_{\mathbb{A}} \pi^{i}(du^{i}\vert y^{i}) \tilde\nu(dy^{i})~~\forall~\mathbb{A}\in \mathcal{B}(\mathbb{U}\times \mathbb{Y})\right\},
    \end{aligned}
    \end{align}
    for $i=1, \dots, N$, equipped with the $w$-$s$ topology. Recall that $\mathbb{U}^{0}, \mathbb{U}^{M}$, and $\mathbb{U}$ are compact and the marginals on observations are independent of $N$. Therefore, the sets of randomized strategies $\Delta^{0, M}(\hat{\nu})$, $\Delta^{M}(\tilde\nu^{M})$, and $\Delta^{i}(\tilde\nu)$ are tight, per \cite[Proof of Theorem 2.4]{yukselSICON2017}. Since these sets are closed, they are compact in the $w$-$s$ topology.
    
    
    By  Proposition \ref{lem:exi-1}(ii), there exists an $(\epsilon^{0}, \hat{\epsilon})$-Stackelberg-incentive equilibrium
    $(\pi^{0\star}_{N}, \pi^{M\star}_{N}, \pi^{\star}_{N}, \dots, \pi^{\star}_{N})$,  symmetric among minor followers, for $\PNtwo$ for each $N$. Let these equilibrium policies induce 
    $(P_{N}^{0, M\star}, P_{N}^{\star}, \ldots, P_{N}^{\star})\in \Delta^{0,M}(\hat{\nu}) \times \prod_{i=1}^{N}\Delta^{i}(\tilde\nu)$. By the compactness of $\Delta^{0, M}(\hat{\nu})$, $\Delta^{M}(\tilde\nu^{M})$, and $\Delta^{i}(\tilde\nu)$, there exist  subsequences  $\{P_{n}^{0,M\star}\}_{n}, \{P_{n}^{\star}\}_{n}$ that converge weakly to $P_{\infty}^{0,M\star}, P_{\infty}^{\star}$ in $\Delta^{0,M}(\hat{\nu}) \times \Delta^{i}(\tilde\nu)$. Let these limiting distributions be induced by $\pi^{0\star}_{\infty}, \pi^{M\star}_{\infty}$, and  $\pi^{\star}_{\infty}$. 
    
    \item [{\it Step 2.}] In this step, we show that $(\pi^{\star}_{\infty}, \mu^{\star}_{\infty})\in R_{\ma, \pi}^{\infty, \hat\epsilon}(\pi^{M\star}_{\infty})$.
    Let $e_{n}\in \mathcal{P}(\mathbb{U}\times \mathbb{Y})$ be the empirical measure given by 
    \begin{flalign}
    e_{n}(\mathbb{A}) := \frac{1}{n}\sum_{i=1}^{n}\delta_{(u^{i\star}_{n}, y^{i})}(\mathbb{A}),
    \end{flalign}
    for any Borel set $\mathbb{A}$ in $\mathbb{U}\times \mathbb{Y}$, where $u^{i\star}_{n}$ is induced by $\pi^{\star}_{n}$. Under Assumption \ref{assum:5}(v), the observations $y^{i}$'s are i.i.d., and hence, $e_n$ converges weakly to $e_{\infty}:=\mathcal{L}(u^{i\star}_{\infty}, y^{i})$, $\mathbb{P}^{0}\times \tilde\nu^{0}\times \tilde\nu^{M}$-a.s., as $n\to \infty$. Thus, a fixed minor follower's expected cost becomes
    \begin{subequations}
    \begin{flalign}
    &\lim_{n\to \infty}J^{i}_{n}(\pi^{M\star}_{n},\pi^{\star}_{n}, \ldots, \pi^{\star}_{n})
    \nonumber
    \\
    &=\lim_{n \to \infty} \int c\left(\omega_{0}, u^{i}, u^{M}, \int u e_{n}(du\times \mathbb{Y})\right)\pi^{M\star}_{n}(du^{M}\vert y^{M})\nonumber\\
    &\qquad\times\prod_{i=1}^{n}\pi^{\star}_{n}(du^{i}\vert y^{i})\mathbb{P}(d\omega_{0}, dy^{M}, dy^{1:n})
    \label{eq:step2.Jn}
    \\
    &=\lim_{n \to \infty} \int c\left(\omega_{0}, u^{i}, u^{M}, \int u e_{n}(du\times \mathbb{Y})\right)P^{M\star}_{n}(du^{M}, dy^{M}) \nonumber\\
    &\qquad \times\prod_{i=1}^{n}P^{\star}_{n}(du^{i}, dy^{i})g_{n}(\omega_{0}, y^{M}
    , y^{1:n})\mathbb{P}^{0}(d\omega_{0})
    \label{eq:step3.2}\\
    &=\int c\left(\omega_{0}, u^{i}, u^{M}, \int u e_{\infty}(du\times \mathbb{Y})\right)P^{M\star}_{\infty}(du^{M}, dy^{M})\nonumber\\
    &\qquad \times \prod_{i=1}^{\infty}P^{\star}_{\infty}(du^{i}, dy^{i})g_{\infty}(\omega_{0}, y^{M}, y^{1:\infty})\mathbb{P}^{0}(d\omega_{0})
    \label{eq:step3.3}
    \\
    &=J^{R}_{\infty}(\pi^{M\star}_{\infty},\pi^{\star}_{\infty}, e_{\infty}),
    \label{eq:step3.4}
    \end{flalign}
    \label{eq:step3.all}
    \end{subequations}
    where  
    \begin{align}
        g_{n}(\omega_{0}, y^{M}, y^{1:n}):=h^{M}(\omega_{0}, y^{M})\prod_{i=1}^{n}h(\omega_{0}, y^{i}), \\
        g_{\infty}(\omega_{0}, y^{M}, y^{1:\infty}):=h^{M}(\omega_{0}, y^{M})\prod_{i=1}^{\infty}h(\omega_{0},y^{i}).
    \end{align}
     Assumption \ref{assum:5}(v) gives $g_{N}$ is bounded. The above derivation makes use of the generalized dominated convergence theorem for varying measures \cite[Theorem 3.5]{serfozo1982convergence} and Fubini's theorem, upon utilizing the boundedness and continuity of $c$ from Assumption \ref{assum:5}.

     Next, we show that  $(\pi^{\star}_{\infty}, \mu^{\star}_{\infty})\in R_{\ma, \pi}^{\infty, \hat\epsilon}(\pi^{M\star}_{\infty})$, given that $(\pi^{\star}_{n}, \ldots, \pi^{\star}_{n})\in R_{\ma, \pi}^{\hat{\epsilon}}(\pi^{M\star}_{n})$. To that end,   $\tilde{e}_{n}\in \mathcal{P}(\mathbb{U}\times \mathbb{Y})$, defined by
    \begin{align}
        \tilde{e}_{n}(\mathbb{A}) = \frac{1}{n}\left[\sum_{i=1, i\not=R}^{n}\delta_{(u^{i\star}_{n}, y^{i})}(\mathbb{A}) + \delta_{(u^{R},y^{R})}(\mathbb{A})\right]
    \end{align} 
    for every Borel set $\mathbb{A}$ in $\mathbb{U}\times \mathbb{Y}$, converges weakly to $e_{\infty}:=\mathcal{L}(u^{i\star}_{\infty},y^{i})$, $\mathbb{P}^{0}\times \tilde\nu^{0}\times \tilde\nu^{M}$-a.s., 
    as $n\to \infty$. For an arbitrary randomized strategy $\pi^{R}$ that induces a distribution on $u^{R}$, we have
    \begin{subequations}
    \begin{flalign}
    &\lim_{n\to \infty}J^{i}_{n}-\hat{\epsilon} \nonumber\\
    &\leq \lim_{n \to \infty} \inf_{P^{R}\in \Delta^{R}(\tilde\nu)} \int c\left(\omega_{0},  u^{R}, u^{M}, \int u \tilde{e}_{n}(du\times \mathbb{Y})\right)P^{M\star}_{n}(du^{M}, dy^{M})\nonumber\\
    & \qquad \times P^{R}(du^{R}, dy^{R}) \prod_{i=1, i\not=R}^{n}P^{\star}_{n}(du^{i}, dy^{i})g_{n}(\omega_{0}, y^{M}, y^{1:n})\mathbb{P}^{0}(d\omega_{0})\label{eq:step3.4bb}\\
    & \leq \inf_{P^{R}\in \Delta^{R}(\tilde\nu)} \int c\left(\omega_{0}, u^{R}, u^{M}, \int u {e}_{\infty}(du\times \mathbb{Y})\right)P^{M\star}_{\infty}(du^{M}, dy^{M})\nonumber\\
    & \qquad \times P^{R}(du^{R}, dy^{R}) \prod_{i=1, i\not=R}^{\infty}P^{\star}_{\infty}(du^{i}, dy^{i})g_{\infty}(\omega_{0}, y^{M}, y^{1:\infty})\mathbb{P}^{0}(d\omega_{0})\label{eq:step3.6}, 
    \end{flalign}
    \label{eq:step3.6.all}
    \end{subequations}
    where \eqref{eq:step3.6.all} follows from an analogous argument as that used in \eqref{eq:step3.all}.
    Then, \eqref{eq:step3.4} implies that $(\pi^{\star}_{\infty}, \mu^{\star}_{\infty})\in R_{\ma, \pi}^{\infty, \hat\epsilon}(\pi^{M\star}_{\infty})$, completing the proof of step 2.
    
    \item [{\it Step 3.}] Proceeding along the same lines as \eqref{eq:step2.Jn}--\eqref{eq:step3.6}, we get that $(\pi^{0\star}_{\infty}, \pi^{M\star}_{\infty}, \pi^{\star}_{\infty})$ constitutes an $(\epsilon^{0}, \hat\epsilon)$-leader-major optimal solution for $\Pitwo$. Details are omitted for brevity.
    
    Similarly, we also get
    \begin{flalign}
    \lim_{n\to \infty}J^{M}_{n}(\pi^{0*}_{n},\pi^{M\star}_{n},\pi^{\star}_{n}, \ldots, \pi^{\star}_{n})= J^{M}_{\infty}(\pi^{0*}_{\infty},\pi^{M\star}_{\infty},\pi^{\star}_{n}, e_{\infty}).\label{eq:step3.117}
    \end{flalign}
    
    To prove that $(\pi^{M\star}_{\infty}, \pi^{\star}_{\infty}, \mu^{\star}_{\infty})\in R_{\pi}^{\infty, \hat\epsilon}(\pi^{0\star}_{\infty})$, define the set of randomized strategies for the leader as 
    \begin{align}
    &\Delta^{0}_{\MCS}(\pi^{M}, \hat{\nu})\nonumber\\
    &:=\bigg\{P \in \mathcal{P}(\mathbb{U}^{0}\times \mathbb{U}^{M}\times \mathbb{Y}^{0} \times \mathbb{U}^{M}) \bigg\vert~\forall \mathbb{A} \in \mathcal{B}(\mathbb{U}^{0}\times \mathbb{U}^{M}\times \mathbb{Y}^{0} \times \mathbb{U}^{M})\nonumber\\
    & \qquad P(\mathbb{A}) = \int_{\mathbb{A}} \pi^{0}(du^{0}\vert y^{0}, y^{M}, u^{M})\pi^{M}(du^{M}\vert y^{M})\hat{\nu}(dy^{0}, dy^{M})\bigg\}.
    \end{align}
     From step 1,  $\Delta^{0}_{\MCS}(\pi^{M}, \hat{\nu})$ is compact in the $w$-$s$ topology, and hence, there exists a converging subsequence $\{P_{n, \pi^{M}}^{0\star}\}_{n}$, which admits the limit $P_{\infty, \pi^{M}}^{0\star}$. Since $(\pi^{0\star}_{n}, \pi^{M\star}_{n}, \pi^{\star}_{n}, \dots,\pi^{\star}_{n})$ is an $(\epsilon_{0}, \hat\epsilon)$-Stackelberg-incentive equilibrium for $\mathcal{P}^{\ma}_{n}$, we have
     \begin{subequations}
    \begin{flalign}
    &\lim_{n\to \infty}J^{M}_{n}(\pi^{0*}_{n},\pi^{M\star}_{n},\pi^{\star}_{n}, \ldots, \pi^{\star}_{n})-\hat{\epsilon} \nonumber\\
    &\leq \lim_{n \to \infty} \inf_{\pi^{M}\in \Delta^{M}(\tilde\nu^{M})}\int c^{M}\left(\omega_{0}, u^{0}, u^{M}, \int u {e}_{n}(du\times \mathbb{Y})\right)\mathbb{P}^{0}(d\omega_{0})\nonumber\\ 
    &\qquad  \times P^{0\star}_{n, \pi^{M}}(du^{0}, dy^{0}, du^{M}, dy^{M})\prod_{i=1}^{n}P_{n}^{\star}(du^{i}, dy^{i})h^{0}(\omega_{0}, y^{0}){g}_{n}(\omega_{0}, y^{M}, y^{1:n})\label{eq:step5.4}\\
    & \leq \inf_{\pi^{M}\in \Delta^{M}(\tilde\nu^{M})} \int c^{M}\left(\omega_{0}, u^{0}, u^{M}, \int u {e}_{\infty}(du\times \mathbb{Y})\right)\mathbb{P}^{0}(d\omega_{0})\nonumber\\
    & \qquad \times P^{0\star}_{\infty, \pi^{M}}(du^{0}, dy^{0}, du^{M}, dy^{M})\prod_{i=1}^{\infty}P_{\infty}^{\star}(du^{i}, dy^{i})h^{0}(\omega_{0}, y^{0}){g}_{\infty}(\omega_{0}, y^{M}, y^{1:\infty})\label{eq:step5.6}\\
    &=\inf_{\pi^{M}\in \Delta^{M}(\tilde{\nu}^{M})} J^{M}_{\infty}(P^{0\star}_{\infty, \pi^{M}},\pi^{M},e_{\infty}),\label{eq:step3.120}
    \end{flalign}
    \label{eq:step3.120.all}
    \end{subequations}
    where \eqref{eq:step3.120.all} follows from an analogous argument as that used in \eqref{eq:step3.all}.
    
    Hence by step 2, we get $(\pi^{M\star}_{\infty}, \pi^{\star}_{\infty}, \mu^{\star}_{\infty})\in R_{\pi}^{\infty, \hat\epsilon}(\pi^{0\star}_{\infty})$. Using the approximate leader-major optimality of $(\pi^{0\star}_{\infty},\pi^{M\star}_{\infty}, \pi^{\star}_{\infty})$, we infer
\begin{flalign}
&\inf_{\pi^{0}\in \Delta_{\MCS}^{0}} \inf_{(\pi^{M}, \pi^{R}, \mu) \in R^{\infty, \hat\epsilon}(\pi^{0})} J^{0}_{\infty}(\pi^{0}, \pi^{M}, \pi^{R})\nonumber\\
&\geq \inf_{(\pi^{0}, \pi^{M}) \in \Delta_{\MCS}^{0}\times \Delta^{M}} \inf_{(\pi^{R}, \mu) \in R^{\infty, \hat\epsilon}_{\ma, \pi}(\pi^{M})} J^{0}_{\infty}(\pi^{0}, \pi^{M}, \pi^{R})\nonumber\\
&\geq J_{\infty}^{0}({\pi}^{0\star}_{\infty}, \pi^{M\star}_{\infty}, \pi^{\star}_{\infty})-\epsilon_{0}\label{eq:Fin13}.
\end{flalign}
This completes the proof of part (i).
\end{itemize}

Part (ii): Since $(\pi^{0\star}, \pi^{M\star}, \pi^{\star})$ constitutes an ${\epsilon}$-mean-field incentive equilibrium for $\Pitwo$, proceeding along the same lines as \eqref{eq:step3.all}--\eqref{eq:step3.6.all}, there exists $\hat\epsilon_{N}$ with $\hat\epsilon_{N}\to \hat\epsilon$ as $N\to \infty$ such that
\begin{flalign}
\left\vert\inf_{\pi^{i} \in \Delta^{i}}  J_{N}^{i}(\pi^{0\star}, \pi^{i}, \pi^{M\star}, \pi^{-i\star})-J_{N}^{i}(\pi^{0\star},\pi^{\star}, \pi^{M\star}, \pi^{\star}, \ldots, \pi^{\star})\right\vert<\hat\epsilon_{N}
\end{flalign}
for all $i=1, \ldots, N$. Similarly, there exists $\epsilon_{N}^{0}$ with $\epsilon_{N}^{0}\to \epsilon^{0}$ as $N\to \infty$ such that
\begin{flalign}
&\bigg\vert\inf_{\pi^{0}, \pi^{M} \times \Delta_{\MCS}^{0}\times \Delta^{M}} \inf_{\pi^{1:N}\in R_{\ma}^{\hat\epsilon_{N}}(\pi^{M})\cap \Delta^{\SYM}} J_{N}^{0}(\pi^{0},\pi^{M}, \pi^{1:N})\nonumber\\
&\qquad \qquad -J_{N}^{0}(\pi^{0\star},\pi^{M\star}, \pi^{\star}, \ldots, \pi^{\star})\bigg\vert<\epsilon^{0}_{N}.
\end{flalign}
Finally, an argument along the same lines as \eqref{eq:step3.117}--\eqref{eq:step3.120.all} gives
\begin{flalign}
\left\vert\inf_{\pi^{M} \in \Delta^{M}}  J_{N}^{M}(\pi^{0\star},\pi^{M}, \pi^{\star}, \ldots, \pi^{\star})-J_{N}^{M}(\pi^{0\star},\pi^{M\star}, \pi^{\star}, \ldots, \pi^{\star})\right\vert<\hat\epsilon_{N},
\end{flalign}
Since $(\pi^{M\star}, \pi^{\star}, \ldots, \pi^{\star})\in R^{\hat\epsilon_{N}}_{\pi}(\pi^{0\star})$ and $(\pi^{0\star},\pi^{M\star}, \pi^{\star}, \ldots, \pi^{\star})$ is an ${\epsilon}_{N}$-leader-major optimal solution for $\PNtwo$, this completes the proof.

\end{document}